\documentclass[twocolumn,preprintnumbers,amsmath,amssymb]{revtex4-1}
\usepackage{epsfig}
\usepackage{color}
\usepackage{amsmath}
\usepackage{float}
\usepackage{amsfonts}
\usepackage{amssymb}
\usepackage{multirow}
\usepackage{graphicx,stackengine}
\usepackage{subfig}
\usepackage[percent]{overpic}
\usepackage[english]{babel}
\usepackage{float}
\usepackage{array}
\usepackage{pst-plot}
\usepackage[version=4]{mhchem}
\usepackage{lipsum}

\usepackage{verbatim}
\usepackage{fancyvrb}
\usepackage{tikz}
\usetikzlibrary{calc,fadings}
\tikzfading[name=fade l,left color=transparent!100,right color=transparent!0]
\tikzfading[name=fade r,right color=transparent!100,left color=transparent!0]
\tikzfading[name=fade d,bottom color=transparent!100,top color=transparent!0]
\tikzfading[name=fade u,top color=transparent!100,bottom color=transparent!0]

\begin{document}
\title{Direct calculation of the planar NaCl-aqueous solution interfacial free energy at the solubility limit}

\author{Ignacio Sanchez-Burgos$^{1}$ and Jorge R. Espinosa$^{1,*}$}
\affiliation{
[1] Maxwell Centre, Cavendish Laboratory, Department of Physics, University of Cambridge, J J Thomson Avenue, Cambridge CB3 0HE, United Kingdom. \\
* = To whom correspondence should be sent.
email: jr752@cam.ac.uk}

\date{\today}

\begin{abstract}

Salty water is the most abundant electrolyte aqueous mixture on Earth, however, very little is known about the NaCl-saturated solution interfacial free energy ($\gamma_s$). Here, we provide the first direct estimation of $\gamma_s$ for several NaCl crystallographic planes by means of the Mold Integration technique, a highly efficient computational method to evaluate interfacial free energies with anisotropic crystal resolution. Making use of the JC-SPC/E model, one of the most benchmarked force fields for NaCl/water solutions, we measure $\gamma_s$ of four different crystal planes, (100), (110), (111), and (11$\overline{2}$) with the saturated solution at normal conditions. We find high anisotropy between the different crystal orientations with values ranging from 100 to 150 mJ m$^{-2}$, \textcolor{black}{and the average value of the distinct planes being} $\overline{\gamma}_s$ = 137(20) mJ m$^{-2}$. This value for the coexistence interfacial free energy is in reasonable agreement with previous extrapolations from nucleation studies. Our work represents a milestone in the computational calculation of interfacial free energies between ionic crystals and aqueous solutions.

\end{abstract}
\maketitle


Electrolyte solutions, and more specifically NaCl solutions, are ubiquitous, with sodium chloride being the major component of sea salt \cite{lyman1940composition} and atmospheric aerosols \cite{poschl2005atmospheric}, as well as playing a key role in atmospheric chemistry \cite{martin2000phase}. At salt concentrations greater than the solubility, crystals of the ionic salt can be formed, giving rise to a liquid-solid interface. The amount of energy per unit of area required to form such interface is known as the interfacial free energy ($\gamma_s$). However, despite the fact that this magnitude is highly relevant in controlling salt precipitation, no experimental techniques have been able to accurately measure $\gamma_s$ for planar liquid-solid interfaces \cite{ickes2015classical,bahadur2007surface}. Therefore, computational techniques can be useful to provide guidance on such important magnitude.

Computational approaches to directly evaluate the liquid-crystal $\gamma_s$ include Cleaving \cite{broughton1986molecular}, tethered Monte Carlo \cite{fernandez2012equilibrium}, Metadynamics \cite{angioletti2010solid}, Mold Integration \cite{espinosa2014mold}, Capillary Wave Fluctuations \cite{hoyt2001method}, and other related thermodynamic integration schemes \cite{benjamin2014crystal,schilling2009computing,bultmann2020computation}. These techniques have been proven to provide reliable estimates of the liquid-solid interfacial free energy for different crystallographic planes and numerous soft matter systems \cite{sanchez2021fcc,espinosa2015crystal,espinosa2016ice,soria2018simulation,benet2015interfacial,davidchack2003direct,davidchack2010hard,ambler2017solid,asta2002calculation,algaba2022simulation}. However, for the case of the NaCl-saturated water solution interface, none of these methods have yet been implemented due to their high computational cost. \textcolor{black}{Currently, the only available estimates of $\gamma_s$ for the NaCl-aqueous solution have been obtained at deep supersaturation via computational nucleation studies using Seeding \cite{lamas2021homogeneous} and Forward Flux Sampling \cite{jiang2018forward}, as well as through experimental measurements of the nucleation rate \cite{na1994cluster}. Then, by means of the Classical Nucleation Theory (CNT) \cite{ZPC_1926_119_277_nolotengo,becker-doring}, $\gamma_s$ has been estimated for curved interfaces under supersaturation conditions.} By extrapolating such results to the saturation concentration, we find the first approximation to the planar NaCl-aqueous saturated solution $\gamma_s$. However, this is not an entirely satisfactory approach given that it relies on the CNT framework, order parameters to identify the number of particles in the clusters \cite{zimmermann2018nacl}, and does not provide any anisotropic crystal information on $\gamma_s$.

In this work, we calculate the interfacial free energy at normal conditions of the NaCl-aqueous solution at the solubility limit for different crystal planes: (100), (110), (111), and (11$\overline{2}$). We choose the SPC/E water model \cite{berendsen1987missing} in combination with the Joung-Cheetham parametrization (JC) for Na$^+$ and Cl$^-$ ions \cite{joung2009molecular} (further details on the force field parameters and simulation details can be found in the Supplementary Material, SM, which includes Refs. \cite{lorentz1881ueber,berthelot1899methode,zeron2022simulation,bekker1993gromacs,bussi2007canonical,parrinello1981polymorphic,hockney1974quiet,darden1993particle,essmann1995smooth,hess1997lincs,espinosa2016seeding,sanchez2022homogeneous,sanchez2021parasitic}) since it can reasonably reproduce the experimental behaviour of NaCl aqueous solutions \cite{joung2008determination,orozco2014molecular,joung2009molecular}. Moreover, this model has been extensively used to benchmark solubility calculations employing different techniques \cite{benavides2016consensus,espinosa2016calculation,mester2015temperature,mouvcka2013molecular}, \textcolor{black}{resulting in a solubility of m=3.71 $mol$·$kg^{-1}$ \cite{nezbeda2016recent}, moderately lower than the experimental one, 6.15 $mol$·$kg^{-1}$. We use the GROMACS Molecular Dynamics package \cite{bekker1993gromacs} in combination with the Mold Integration (MI) technique \cite{espinosa2014mold}, where formation of a solid slab in the solution} is performed along a reversible pathway, and the free energy difference between the initial (aqueous solution) and final states (aqueous solution + crystal slab) corresponds to $\gamma_s$ times the area of the induced liquid-solid interface. Through MI, $\gamma_s$ can be obtained as:

\begin{figure*}

        
  
    
\includegraphics[width=\linewidth]{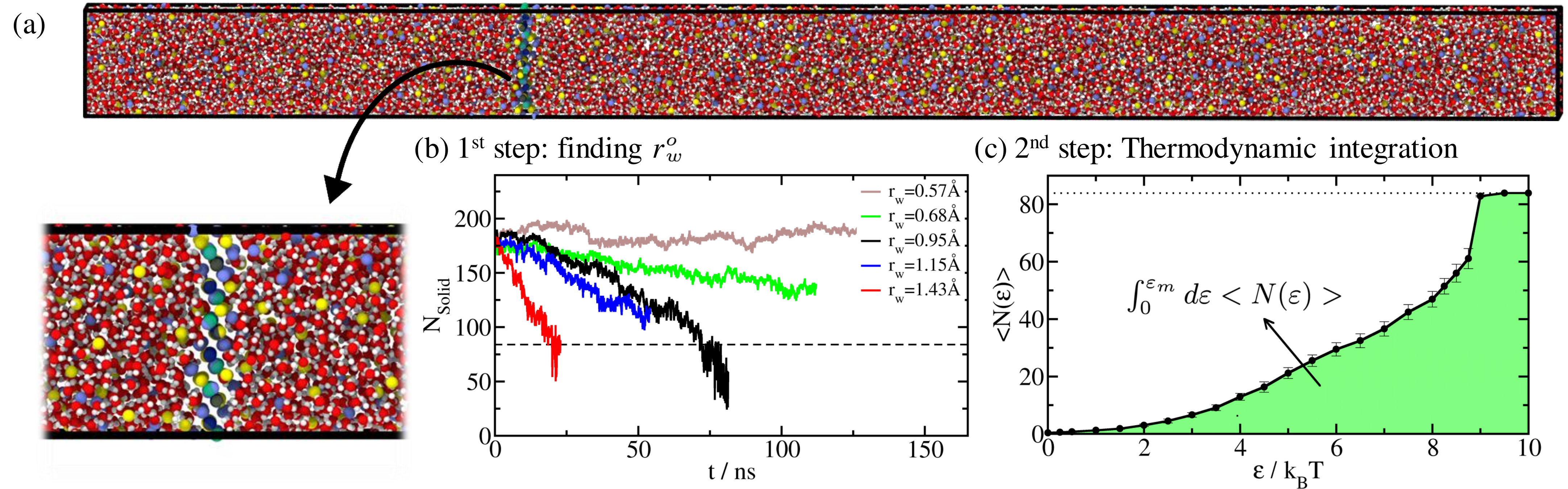}
\caption{Determination of $\gamma_s$ for the (11$\overline{2}$) crystal plane: (a) Representative simulation box employed for the MI technique, along with a close view of the mold occupied by ions. The images were rendered using OVITO \cite{stukowski2009visualization}. (b) First step of the MI calculation to determine the optimal well radius ($r_w^o$): Time-evolution of the number of ions conforming the NaCl crystal slab is depicted for different well widths ($r_w$). The horizontal dashed line indicates the total number of ions that can be accommodated within the mold. N$_{solid}$ was determined through the $\bar{q}_4$-$\bar{q}_6$ local order parameter \cite{lechner2008accurate} (further details in the SM). (c) Second step of MI calculations: simulations at different well depth ($\varepsilon$) values for a fixed $r_w$ are performed to evaluate the integral from Eq. \ref{eqgamma}. \textcolor{black}{The average number of occupied wells ($<N(\varepsilon)>$) against $\varepsilon$ is plotted here for $r_w$=0.78 \r{A}. The green shaded area gives the integral of Eq.\ref{eqgamma}.}}
    \label{figure1}
\end{figure*}

\begin{equation}
    \gamma_s=\frac{1}{2A}\left( \varepsilon_m N_w -\int_0^{\varepsilon_m}d\varepsilon<N(\varepsilon)>\right)
    \label{eqgamma}
\end{equation}

where $\varepsilon$ is the energy of the potential wells (and $\varepsilon_m$ the maximum depth employed), $A$ is the surface of the liquid-solid interface, $N_w$ the number of wells in the mold, and $<N(\varepsilon)>$ the average number of occupied wells at a given potential energy depth value. \textcolor{black}{The method consists on performing thermodynamic integration (TI) along the path in which the depth of the mold potential wells is gradually increased to a maximum value of $\varepsilon_m$. To ensure reversibility in Eq. \ref{eqgamma}, the crystal structure induced by the mold must quickly melt when the interaction between the potential wells and the fluid is switched off. Consequently, the TI has to be performed at well radii ($r_w$) that are wider than the optimal one, $r^o_w$, at which the crystal slab is fully formed, and therefore can possibly induce irreversible crystal growth (i.e., leading to an overestimation of $<N(\varepsilon)>$). Therefore, $\gamma_s(r_w)$ is estimated for several values of $r_w > r^o_w$, and then, extrapolated to $r^o_w$, which is the well radius that recovers the exact free energy value $\gamma_s$ \cite{sanchez2021fcc,espinosa2015crystal,espinosa2016ice,espinosa2014mold,soria2018simulation}}. In practice, the method consists of two distinct steps \cite{espinosa2014mold}: In the first one, we find $r_w^o$ by performing several simulations in which we identify the largest well width at which the solid layer grows or keeps stable without melting (i.e., $r_w < r^o_w$). In the second step, multiple simulations at radii wider than the optimal one are performed, \textcolor{black}{and we measure the average number of wells occupied for each radius as a function of $\varepsilon$ to solve Eq. \ref{eqgamma}.}

\begin{figure}[]
    
        
\includegraphics[width=\linewidth]{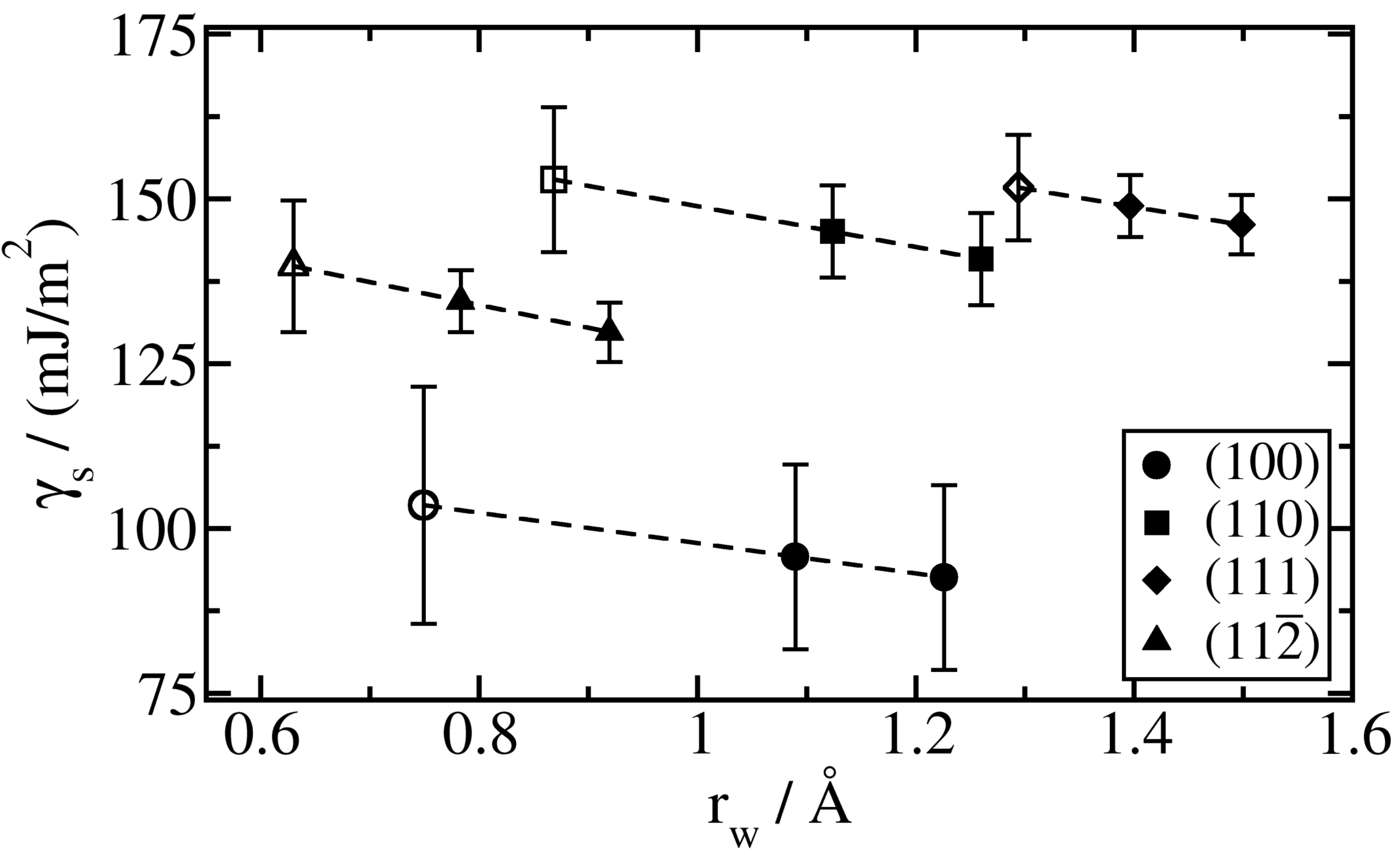}

    \caption{(a) Interfacial free energy as a function of the potential well radius evaluated for four different crystal orientations. Filled symbols indicate $\gamma_s$ obtained through Eq. \ref{eqgamma} for $r_w$ $>$ $r_w^o$, while dashed lines depict linear extrapolations to the optimal well radius $r_w^o$ (empty symbols).}
    \label{figure2}
\end{figure}

In Fig. \ref{figure1}, \textcolor{black}{we describe such procedure for the (11$\overline{2}$) crystal plane at T=298K and p=1 bar}. First, to determine $r_w^o$, we develop a configuration in which the NaCl crystal positions of the mold are already occupied with their corresponding type of ions (Figure \ref{figure1}(a)). \textcolor{black}{Additionally, a crystal layer displaying vacancies (randomly located) in half of the Na$^+$/Cl$^-$ lattice positions is placed at each side of the inserted mold. Importantly, the ions within such adjacent two semioccupied crystal layers are not held through potential wells to retain their equilibrium lattice positions. Since the crystal growth of the NaCl solid at solubility conditions is extremely slow \cite{espinosa2016calculation}, especially for crystal planes with low Miller indices (such as the (100) \cite{kolafa2016solubility}), we can estimate $r_w^o$ in the limit at which each of the adjacent half layers of NaCl ions to the potential mold dissolves or not. If they melt, the potential wells are too wide to induce crystallization, whereas if the ions of the layers aside the mold remain crystalline over long timescales (or even grow), such value of $r_w$ is considered below the optimal radius.} Importantly, to ensure that the solution concentration remains constant at $m\sim$3.7 $mol$·$kg^{-1}$ despite partial melting or growth from the crystal slab (Fig. \ref{figure1}b), we employ system sizes with over 10000 water molecules, which can absorb small variations of ions from the slab to the solution or vice versa (Fig. \ref{figure1}(a)). We use the isothermal-isobaric ($NpT$) ensemble---where pressure is only applied to the perpendicular axis to the crystal-liquid interface---to keep constant both temperature and pressure. \textcolor{black}{We note that an alternative approach to keep constant the solution concentration in our MD simulations without requiring an elongated box is through the grand canonical ensemble \cite{perego2015molecular,karmakar2018cannibalistic}. Nevertheless, it may have only sped up our simulations by a factor of 3 since system sizes of at least 3500 molecules would have been still needed to prevent finite size effects in our MI calculations \cite{espinosa2014mold,sanchez2021fcc}.}

In Figure \ref{figure1}(b), we show the time-evolution of solid-like ions (evaluated through the $\bar{q}_4$-$\bar{q}_6$ local order parameter \cite{lechner2008accurate}; further details on the SM) for different well widths of the (11$\overline{2}$) plane. Here, well radii greater than 0.68 \r{A} results in gradual dissolution of the crystal layers located at each side of the crystal plane induced by the mold. However, for $r_w$=0.57 \r{A} those ions remain ordered aside the crystal slab and even mildly grow over time, hence indicating that such $r_w$ is lower than $r^o_w$. Therefore, we determine $r_w^o$ at the intermediate value of 0.625 \r{A}.

Once $r_w^o$ has been determined, we perform TI to compute the required free energy to induce the formation of the crystal slab. TI requires performing simulations at different well depths ($\varepsilon$) for a fixed $r_w$ and measuring the average occupation of the mold at each $\varepsilon$, which is the integrand of Eq. \ref{eqgamma}. To minimize the extent of irreversibility (due to crystal growth) in these calculations, we integrate at $r_w$ values of 0.78 and 0.92 \r{A}. In Figure \ref{figure1}(c) we show the average number of occupied wells as a function of the well depth ($\varepsilon$) for $r_w$=0.78 \r{A}, where the shaded area corresponds to the integral in Equation \ref{eqgamma}, from which we can directly obtain the interfacial free energy. \textcolor{black}{In Section $SII$ of the SM we include a detailed discussion of the different sources of uncertainty along the integration pathway, including the small hysteresis associated to the steep change in $<N(\varepsilon)>$ at the mold high occupation regime (Fig. S2)}. Once $\gamma_s$ is evaluated for different $r_w$ values, it can be extrapolated to the optimal $r_w^o$.
In Figure \ref{figure2} we show the obtained interfacial free energy for $r_w>r_w^o$ depicted with filled symbols, along with the corresponding extrapolations to the optimal radius, represented with empty symbols. Apart from the (11$\overline{2}$) plane, 
we also evaluate $\gamma_s$ for the (100), (110) and (111) planes. For all planes we follow the same procedure described for the (11$\overline{2}$) face. The final interfacial free energies for the different planes are reported in Table \ref{table1}, where we also include the planar density as the number of ions per nm$^2$, as well as the total number of potential wells employed for the calculation of each crystal plane. For all the different orientations, we make use of 2 layers of potential wells to induce the formation of the crystal slab. For reproducibility purposes, in the SM we provide Source Data links to all the liquid-crystal and pure NaCl solid configurations employed in our MI calculations, \textcolor{black}{along with snapshots of the four planes studied (Fig. S3)}.

\begin{table}[]
    \centering
        \begin{tabular}{c|c|c|c}
         Crystal & Layer density / & \multirow{2}{*}{N$_w$} & \multirow{2}{*}{$\gamma_s$ / (mJ m$^{-2}$)} \\
         plane & (ions nm$^{-2}$) & & \\ \hline
         (100) & 11.967 & 200 & 104 \textcolor{black}{$\pm$} 18 \\
         (110) & 8.462 & 140 & 153 \textcolor{black}{$\pm$} 11 \\
         (111) & 6.909 & 112 & 152 \textcolor{black}{$\pm$} 8 \\
         (11$\overline{2}$) & 4.885 & 84 & 140 \textcolor{black}{$\pm$} 10 \\ \hline
         \multicolumn{4}{c}{\textbf{Average $\overline{\gamma}_s$ = 137 $\pm$ 20 mJ m$^{-2}$}}

        \end{tabular}
    \caption{Values of the ion density per layer, number of potential wells (N$_w$) employed in the MI calculations, and the resulting liquid-solid interfacial free energy ($\gamma_s$) for each of the studied crystal orientation. \textcolor{black}{$\overline{\gamma}_s$ represents the average of the different crystal orientations.}}
    \label{table1}
\end{table}

Strikingly, when comparing the interfacial free energy of the distinct crystal orientations (Fig. \ref{figure2}), we find large differences, of up to 50\% higher values, for the (110) and (111) planes compared to the (100) face (Table \ref{table1}); similarly to the crystal-molten NaCl (although with the Tosi-Fumi model \cite{espinosa2015crystal}).
However, while the differences in $\gamma_s$ between these distinct planes in crystal-molten NaCl were of the order of 5-15 mJ m$^{-2}$ \cite{espinosa2015crystal,benet2015interfacial}, in NaCl-aqueous solutions can reach up to 40-50 mJ m$^{-2}$. \textcolor{black}{We note that within the uncertainty of our calculations (Table \ref{table1}), the anisotropy in $\gamma_s$ is only statistically significant between the (100) plane and the (110) and (111) crystal orientations}. The higher interfacial anisotropy in NaCl-aqueous solutions is also consistent with the fact that the average $\gamma_s$ for the studied planes ($\overline{\gamma}_s$) is $\sim$137 mJ m$^{-2}$, whereas for the crystal-molten NaCl is between 90-100 mJ m$^{-2}$ \cite{espinosa2015crystal,benet2015interfacial}. This is a reasonable result given that in the crystal-molten interface, both phases are formed by particles of the same nature (Na$^+$ and Cl$^-$ ions) and, therefore, the energetic cost to form an interface should be lower \cite{zimmermann2015nucleation}. Nonetheless, the crystal-molten NaCl calculations were performed for the Tosi-Fumi model at its coexistence temperature (1082K), and therefore, this cannot be taken as a direct comparison.

\textcolor{black}{By applying a Wulff's construction \cite{wulff1901xxv,rahm2020wulffpack} (further details provided in the SM), we also determine the shape of the macroscopic NaCl crystals through our calculations, and estimate an average value of the interfacial free energy for such crystals ($\gamma_{s,W}$= 109 mJ m$^{-2}$). The lower average value of $\gamma_{s,W}$ obtained via the Wulff's construction compared to $\overline{\gamma}_s$ can be explained through the much greater contribution of the (100) plane to the macroscopic crystal compared to the rest of crystal orientations studied here. Moreover, in reasonable agreement with experiments \cite{aquilano20091}, the predicted shape by the JC-SPC/E model for the macroscopic NaCl crystal is roughly cubic with the corners cut out by the exposure of the (11$\overline{2}$) plane (Fig. S4).}

Interestingly, we also note that there is no clear correlation between the plane density and interfacial free energy of the studied orientations (Table \ref{table1}), in contrast to some previously investigated systems such as Hard-Spheres \cite{mu2005anisotropic,davidchack2010hard,benjamin2015crystal,schmitz2015ensemble} or Lennard-Jones \cite{laird2009determination,davidchack2003direct,espinosa2014mold}. The reason behind such observation in Hard-Spheres or Lennard-Jones systems is that higher planar density usually implies higher differences in density between the lower density coexisting liquid and the higher density crystal phase. However, in NaCl-aqueous solutions, although such behaviour also applies, the delicate balance between electrostatic repulsion and ion ordering might additionally modulate $\gamma_s$. 

\begin{figure}
    \centering
    \includegraphics[width=\linewidth]{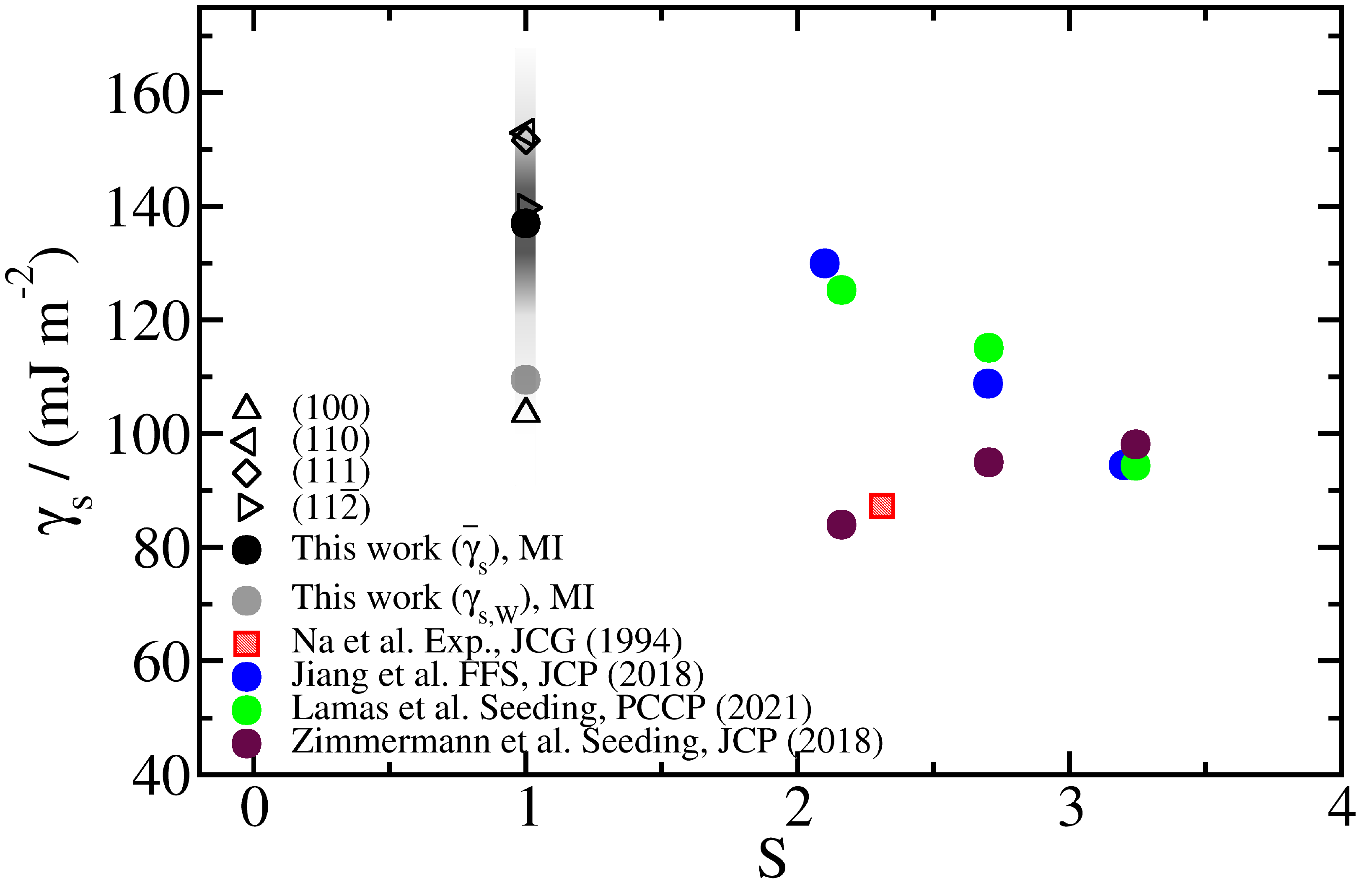}
    \caption{Interfacial free energy ($\gamma_s$) as a function of supersaturation (S=$m/m_{sat}$), being $m_{sat}$=3.71 $mol$·$kg^{-1}$ for the JC-SPC/E model. \textcolor{black}{Our calculations at S=1 for different crystal orientations (empty triangles) as well as for $\overline{\gamma}_s$ and $\gamma_{s,W}$ (filled circles) are depicted by black and grey symbols respectively.} Interfacial free energies obtained from nucleation studies at high supersaturations, both computational \cite{jiang2018forward,lamas2021homogeneous,zimmermann2018nacl} and experimental \cite{na1994cluster}, are also included.}
    \label{figure3}
\end{figure}

We compare our values of $\gamma_s$ at the solubility concentration with those previously estimated from nucleation studies at high supersaturation. From both experimental \cite{na1994cluster} and computational \cite{lamas2021homogeneous,jiang2018forward,zimmermann2018nacl} nucleation rates, an average of $\gamma_s$ (for a curved interface containing contributions of all the possible crystal orientations) can be inferred by means of the Classical Nucleation Theory \cite{ZPC_1926_119_277_nolotengo,becker-doring}. Importantly, since most of the previous computational nucleation studies were performed using the JC-SPC/E model \cite{lamas2021homogeneous,jiang2018forward,zimmermann2018nacl}, we can establish a direct comparison of our results to those from supersaturated concentrations. In Figure \ref{figure3}, we plot the interfacial free energy as a function of supersaturation.
\textcolor{black}{Our results for $\gamma_s$ at coexistence are shown for each of the crystal orientation that we studied (empty triangles) together with the mean value of them ($\overline{\gamma}_s$; black circle) and the average value from the Wulff's construction for the equilibrium crystal ($\gamma_{s,W}$; grey circle).} \textcolor{black}{As can be seen, the extrapolated interfacial free energy trend to S=1 from Lamas \emph{et al.} \cite{lamas2021homogeneous} and Jiang \emph{et al.} \cite{jiang2018forward} are in excellent agreement with our direct calculations of $\gamma_s$ for different crystal planes. However, a significant better agreement is found between the extrapolated interfacial free energy from these nucleation studies and $\overline{\gamma}_s$ (arithmetic mean) than with the obtained $\gamma_{s,W}$ from the Wulff's construction (Fig. \ref{figure3}). That might be explained by the fact that in nucleation studies the typical size of the NaCl clusters is of the order of tens of ions (i.e., from 10 to 100 ions \cite{lamas2021homogeneous,jiang2018forward,zimmermann2018nacl}), and their shape is roughly spherical.
Hence, the overall $\gamma_s$ for these small critical nuclei may be contributed by several distinct crystallographic planes, interfacial defects, curvature effects, or by the Laplace pressure \cite{montero2020interfacial,espinosa2016calculation}. In fact, even large critical nuclei stable at much less supersaturated concentrations (i.e., S$\sim$1.5) typically display spherical shapes with curvature effects \cite{espinosa2016calculation}. In contrast, at the saturation concentration, macroscopic roughly cubic crystals mainly exposing the (100) plane (with a possible small contribution of the (11$\overline{2}$), or (111) planes on the vertices \cite{aquilano20091}) are expected to be formed displaying an overall interfacial free energy that highly resembles to that of the (100) plane: $\gamma_{s,W}$= 109 mJ m$^{-2}$ \textit{vs.} $\gamma_{s,(100)}$= 104 mJ m$^{-2}$. Such mostly cubic shape of the equilibrium NaCl crystal predicted through the Wulff's construction (Fig. S4) is in good agreement with experimental observations for macroscopic NaCl crystallites \cite{quilaqueo2016crystallization,aquilano20091}.}

On the contrary, the extrapolated trend from Zimmermann \emph{et al.} \cite{zimmermann2018nacl} significantly underestimates $\overline{\gamma}_s$ and $\gamma_{s,W}$ at coexistence conditions (Fig. \ref{figure3}). That is not surprising considering that the nucleation rates from which the interfacial free energies were obtained in Ref. \cite{zimmermann2018nacl} severely overestimated those from Refs. \cite{lamas2021homogeneous,jiang2018forward}. Importantly, the $\gamma_s$ dependence with supersaturation which reasonably extrapolates to our calculations (those from Refs. \cite{lamas2021homogeneous,jiang2018forward}) suggests that $\gamma_s$ decreases as the salt concentration increases. This observation would be consistent with the fact that the chemical composition of both phases becomes more similar with supersaturation, and thus, at the limit of infinite supersaturation (molten NaCl), the interfacial free energy should be lower than at the solubility limit \cite{espinosa2015crystal}. The observed substantial differences in $\gamma_s$ from nucleation studies also evidence the critical relevance of the employed local order parameter for determining the nucleus size, and thus, the interfacial free energy \cite{zimmermann2018nacl}. 
Finally, we also compare with the experimental interfacial free energy inferred by Na \emph{et al.} \cite{na1994cluster} using the CNT framework (Fig. \ref{figure3}, red square), which is significantly below the predicted $\gamma_s$ from Refs. \cite{jiang2018forward,lamas2021homogeneous} (not from Ref. \cite{zimmermann2018nacl}), although it is qualitatively consistent with the hypothesis that the interfacial free energy may decrease with supersaturation. A simple possible explanation for the observed discrepancies between these computational \textit{vs.} experimental nucleation estimates of $\gamma_s$ may be the force field performance, nevertheless, the difficult determination of the experimental CNT kinetic pre-factor to infer the nucleation free energy barrier, from which the interfacial free energy is extracted, might be also a significant source of uncertainty.

In summary, we provide here the first direct measurement of the NaCl-brine solution interfacial free energy at the saturation concentration and normal conditions. 
\textcolor{black}{We overcome technical difficulties of these calculations, such as the slow crystal growth dynamics, by employing the Mold Integration}, a computational technique which evaluates the free energy work to form a crystal slab from the saturated solution. By using the JC-SPC/E model, one of the most benchmarked force fields for NaCl in water, we measure the interfacial free energy of four different planes: the (100), (110), (111), and (11$\overline{2}$); obtaining an average value of $\overline{\gamma}_s$ = 137(20) mJ m$^{-2}$. Remarkably, large differences of up to 50 mJ m$^{-2}$ in $\gamma_s$ between the distinct crystal orientations are found.
Finally, we note that our results of $\gamma_s$ at the solubility limit are consistent with extrapolated values from nucleation studies (using the same model) as well as with experimental data inferred from a CNT analysis at high supersaturation. Taken together, this work represents a milestone in the computational calculation of interfacial free energies between aqueous solutions and ionic crystals. \\

This project has received funding from the Oppenheimer Research Fellowship of the University of Cambridge. I.~S.-B. acknowledges funding from Derek Brewer scholarship of Emmanuel College and EPSRC Doctoral Training Programme studentship, number EP/T517847/1. J.~R.~E. also acknowledges funding from the Roger Ekins Research Fellowship of Emmanuel College. This work has been performed using 3 million of CPU hours provided by the Cambridge Tier-2 system operated by the University of Cambridge Research Computing Service (http://www.hpc.cam.ac.uk) funded by EPSRC Tier-2 capital grant EP/P020259/1. We thank V. Roser for critical reading of the manuscript.

\clearpage



\counterwithin{equation}{section}
\counterwithin{table}{section}
\renewcommand\thesection{S\Roman{section}}   
\renewcommand\thefigure{S\arabic{figure}} 
\renewcommand\thetable{S\arabic{table}} 
\renewcommand\theequation{S\arabic{equation}}    
\onecolumngrid
\setcounter{page}{1}
\begin{center}
    {\large \textbf{Supplementary Material: Direct calculation of the planar NaCl-aqueous solution interfacial free energy at the solubility limit } \\ \par} \vspace{0.3cm}
    Ignacio Sanchez-Burgos$^{1}$ and Jorge R. Espinosa$^{1,*}$ \\ 
    $[1]$ Maxwell Centre, Cavendish Laboratory, Department of Physics, \\  University of Cambridge, J J Thomson Avenue, Cambridge CB3 0HE, United Kingdom. \\
* = To whom correspondence should be sent.
email: jr752@cam.ac.uk

\end{center}
\thispagestyle{empty}
\section{Models and simulation details}

\subsection{Joung-Cheatham-SPC/E model}

The Single Point Charge/Extended (SPC/E) water model \cite{berendsen1987missingkk} employed in this study defines water as a 3-site rigid molecule (1 oxygen and 2 hydrogen atoms), where the O-H distance is fixed at 1 \r{A}, and the H-O-H angle at 109.47$^o$. Furthermore, we use an extension of this water model describing alkali and halide monovalent ions proposed by Joung and Cheetham \cite{joung2009molecularkk}, which includes a parametrization for Na$^+$ and Cl$^-$ ions. Within this force field, ions are represented as single point particles. The intermolecular interactions in the Joung-Cheeatham-SPC/E (JC-SPC/E) model are defined by a combination of Lennard-Jones and Coulombic potentials:

\begin{equation}
    U_{JC-SPC/E}=4\varepsilon_{ij}\left[ \left(\ \frac{\sigma_{ij}}{r_{ij}} \right)^{12} - \left(\ \frac{\sigma_{ij}}{r_{ij}} \right)^{6} \right] + \frac{1}{4\pi\varepsilon_0} \frac{q_{i}q_j}{r_{ij}}
\end{equation}

where $i$ and $j$ are the different atoms (O, H, Na$^+$ and Cl$^-$), $\varepsilon_{ij}$ is the depth of the Lennard Jones potential between particles $i$ and $j$, $\sigma_{ij}$ is the effective molecular diameter between particles $i$ and $j$, r$_{ij}$ is the distance that separates particles $i$ and $j$ at the moment of evaluating U$_{JC-SPC/E}$, q$_i$ and q$_j$ represent the charges of the particles $i$ and $j$ respectively, and $\varepsilon_0$ is the permittivity of vacuum. For each type of atom, a value of $\varepsilon_i$ and $\sigma_i$ is defined in Table \ref{table1}. The different values $\varepsilon_{ij}$ and $\sigma_{ij}$ can be calculated employing the Lorentz-Berthelot \cite{lorentz1881ueberkk,berthelot1899methodekk} mixing rules so that $\sigma_{ij}=(\sigma_i+\sigma_j)/2$ and $\varepsilon_{ij}=\sqrt{\varepsilon_i\varepsilon_j}$. Further details on this model can be found in Refs. \cite{berendsen1987missingkk,joung2009molecularkk}.

\newcommand{\ncomps}{\widetilde{m}}
\newcommand{\niter}{n_{\text{iter}}}
\newcommand{\cstar}{c^\star}

\begin{table}[h]

\centering

    \begin{tabular}{c|c|c|c|c}
        i & m$_i$ / (g mol$^{-1}$) & q$_{i}$ / q$_H$ & $\varepsilon_i$ / (kJ mol$^{-1}$) & $\sigma_i$ / \r{A} \\ \hline
        O & 15.9994 & -0.8476 & 0.650 & 3.166 \\
        H & 1.008 & 0.4238 & 0 & 0 \\
        Na$^+$ & 22.99 & 1.0 & 1.475 & 2.166 \\
        Cl$^-$ & 35.453 & -1.0 & 0.054 & 4.830 \\
    \end{tabular}

\caption{Mass (m$_i$), charge (q$_i$), $\varepsilon_i$ and $\sigma_i$ values for the different atoms/ions in the JC-SPC/E NaCl-water model.}
\label{table1}
\end{table}

\subsection{Square well potential for the Mold Integration technique}

\begin{figure}
    \centering
    \includegraphics[width=0.55\linewidth]{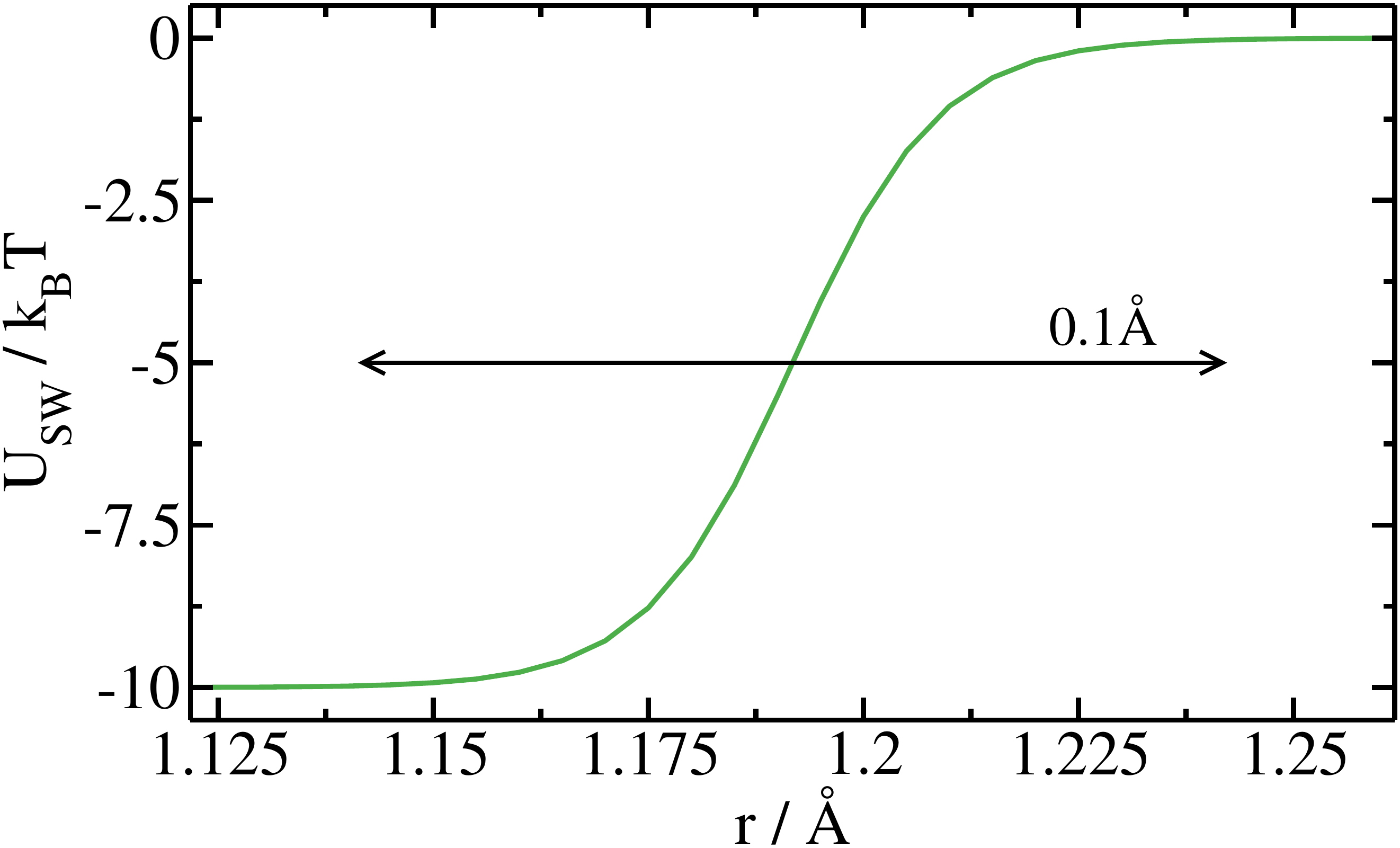}
    \caption{\textcolor{black}{Representation of the continuous square-well potential ($U_{SW}$) as a function of distance from the center of the well ($r$) for a well depth of $\varepsilon_{SW}$= 10k$_B$T, and a well width of $r_w$= 1.19\r{A}. Please note that a zoom in on the transition region from the maximum well depth to the region in which the potential completely vanishes has been applied.}}
    \label{pozo}
\end{figure}

Within the Mold Integration (MI) technique, the formation of the NaCl crystal planes is induced by inserting molds, which consist of multiple potential wells with the perfect NaCl solid structure. The plane exposed to the NaCl aqueous solution is the one for which we obtain the interfacial free energy ($\gamma_s$). The interaction between the wells and the ions is modeled through a square-well like (i.e continuous) potential (SW) \cite{espinosa2014moldkk} of the following form:

\begin{equation}
U_{SW}=-  \frac{1}{2} \varepsilon_{SW} \left[1-\tanh \left( \frac{r-r_w}{\alpha} \right) \right]
\label{poci}
\end{equation}

where $\varepsilon_{SW}$ is the depth of the potential energy well, $r_w$ the radius of the attractive well, $\alpha$ controls the steepness of the well, and $r$ is the distance between the ions and the wells. \textcolor{black}{We choose $\alpha=0.017$ \r{A} as discussed in Refs. \cite{espinosa2014moldkk,espinosa2015crystalkk,espinosa2016icekk,sanchez2021fcckk,zeron2022simulationkk}. Such value of $\alpha$ enables a smooth transition of the potential (of the order of 0.1 \r{A}) from its maximum well depth to the region where it completely vanishes (see Fig. \ref{pozo}).} Further details on this potential can be found in the original reference of the MI method \cite{espinosa2014moldkk}. Within our calculations, specific potential wells for each type of ion in the crystal lattice are employed following Ref. \cite{espinosa2015crystalkk}. \\

\textcolor{black}{To ensure that a correct integration of the equations of motion is taking place with the chosen timestep (2 fs) and steepness of the well (determined by $\alpha = 0.017$ \r{A}), we perform integration points (at 3 and 7.5 k$_B$T) using different timesteps for the integration of the equations of motion. In Table \ref{taboc}, we show the average occupation of the mold ($<$N$(\varepsilon _{SW})>$) with the different employed timesteps. We find that the average value of N$(\varepsilon _{SW})$ is not affected by the choice of the timestep up to simulation timesteps of 5 fs, where values of $<$N$(\varepsilon _{SW})>$ deviate outside of the error (at 7.5k$_B$T) compared to lower values for the integration timestep (i.e., from 0.2 to 2 fs). We note that all points were simulated for a total of 50 ns. Therefore, despite we conclude that a choice of 2 fs ensures accurate results when performing simulations in the $NVT$ or $NpT$ ensemble while providing the fastest computational performance, we recommend for less demanding computational studies making use of a lower timestep (i.e., 1 fs) or moderately increasing the value of $\alpha$.}

\begin{table}[]
    \centering
    \begin{tabular}{>{\color{black}}c >{\color{black}}c | >{\color{black}}c >{\color{black}}c}
    
    \multicolumn{2}{c|}{\textcolor{black}{$\varepsilon _{SW}$=3k$_B$T}} & \multicolumn{2}{c}{\textcolor{black}{$\varepsilon _{SW}$=7.5k$_B$T}} \\ \hline
        Timestep (fs) & $<$N$(\varepsilon _{SW})>$ & Timestep (fs) & $<$N$(\varepsilon _{SW})>$ \\ \hline
         0.2 & 6.6 $\pm$ 2.2 & 0.2 & 42.2 $\pm$ 1.4 \\
         0.5 & 6.6 $\pm$ 2.3 & 0.5 & 41.9 $\pm$ 2.4 \\
         1 & 6.3 $\pm$ 2.2 & 1 & 44.2 $\pm$ 1.9 \\
         2 & 6.3 $\pm$ 2.2 & 2 & 42.4 $\pm$ 2.3 \\
         5 & 6.4 $\pm$ 2.2 & 5 & 30.8 $\pm$ 3.3 \\
    \end{tabular}
    \caption{\textcolor{black}{Average mold occupation ($<$N$(\varepsilon _{SW})>$) for different timesteps and well depths for the (11$\overline{2}$) crystal plane using values of $r_w$ = 0.78 \r{A} and $\alpha$ = 0.017 \r{A}.}}
    \label{taboc}
\end{table}

\subsection{Simulation details}

Simulations are performed with the GROMACS 4.6.7 Molecular Dynamics package \cite{bekker1993gromacskk} in the NpT ensemble (with N equal to the number of particles, p the pressure and T the temperature), keeping T constant at 298.15K with the V-rescale thermostat \cite{bussi2007canonicalkk} and pressure constant at 1bar with the Parrinello-Rahman barostat \cite{parrinello1981polymorphickk}. Pressure is only applied in the long axis of the simulation box (i.e., perpendicular to the liquid-solid interface) using an anisotropic barostat. 
We integrate the equations of motion using the Leap-Frog integrator \cite{hockney1974quietkk}. The simulation timestep chosen is 2 fs, \textcolor{black}{and the thermostat and barostat relaxation times are 1 and 2 ps, respectively}. We set the cut-off of both dispersive interactions and the real part of the electrostatic interactions at 12 \r{A}. Moreover, long-range Coulombic interactions are treated with the Particle-Mesh Ewald (PME) solver in GROMACS \cite{darden1993particlekk,essmann1995smoothkk}. We keep the O-H bond length (1 \r{A}) and H-O-H angle (109.47$^o$) values constant with the LINCS algorithm implemented in GROMACS \cite{hess1997lincskk}. The positions of the wells are kept constant using the \emph{freezegrps} utility in GROMACS. Long range dispersion corrections for energy and pressure are not applied. Nevertheless, we have checked that both the crystal and solution densities at coexistence (i.e., saturation concentration; 3.7 $m$) are consistent with those reported from previous studies (i.e., differing in less than 1\%) using the same model with shorter cut-off distances (e.g., 9 \r{A}) for the potential terms, and long range dispersion corrections \cite{lamas2021homogeneouskk,espinosa2016calculationkk,jiang2018forwardkk,zimmermann2018naclkk}. 

\section{Determination of the uncertainty in MI calculations}

\begin{figure}
    \centering
    \includegraphics[width=0.55\linewidth]{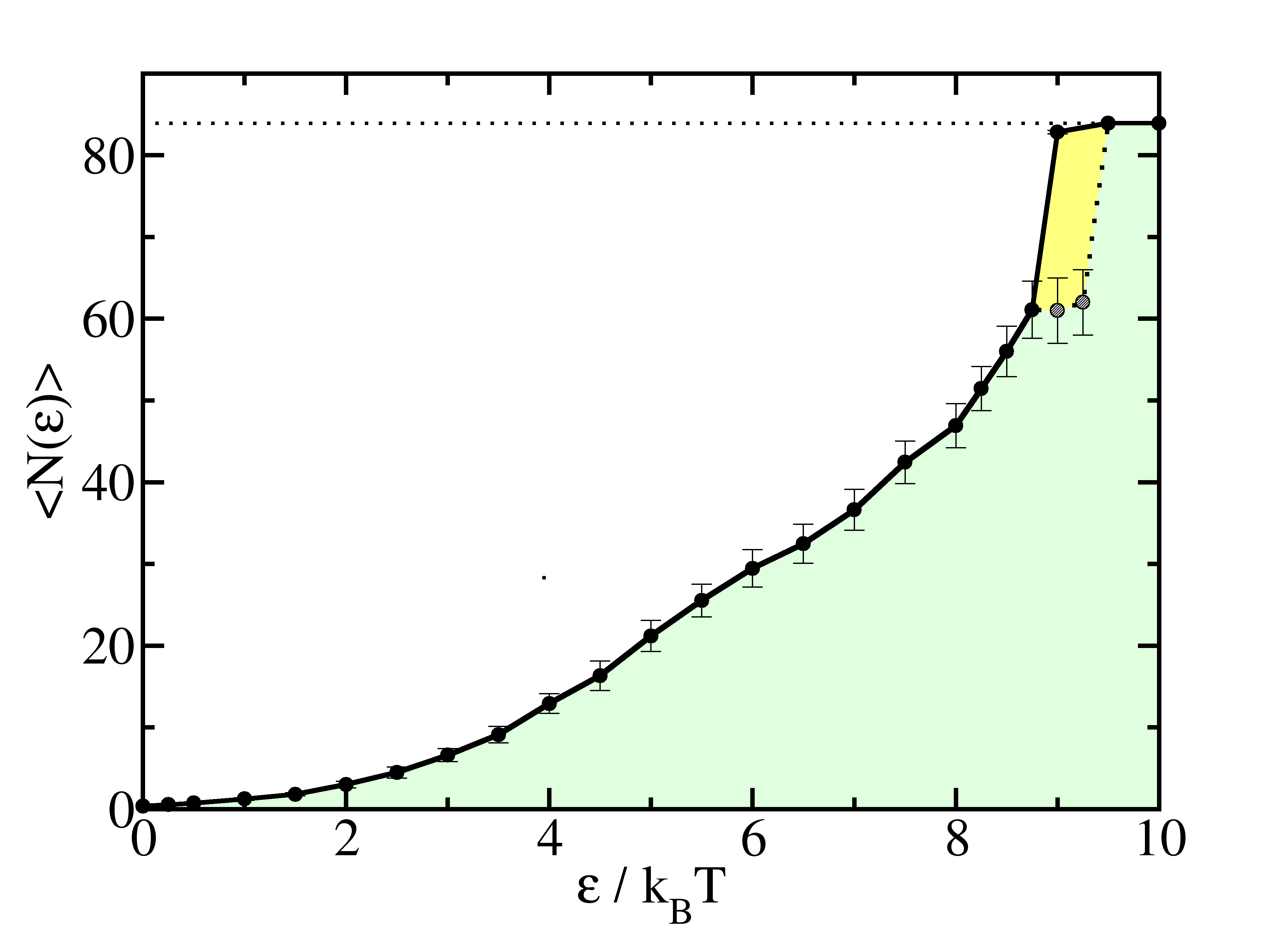}
    \caption{\textcolor{black}{Average number of occupied wells ($<N(\varepsilon)>$) as a function of the potential well depth ($\varepsilon$) for the (11$\overline{2}$) crystal orientation at $r_w$=0.78 \r{A}. Black circles represent values of $<N(\varepsilon)>$ from simulations in which  all the potential wells in the mold were filled at the initial configuration, whereas grey circles those from an starting configuration in which the mold occupancy was relatively low (i.e., $\sim$20\%). The yellow shaded area represents the uncertainty across the integration pathway resulting from hysteresis. Please note that black and grey symbols overlap for the whole integration pathway apart from those shown below the yellow shaded region.}}
    \label{hist}
\end{figure}

\textcolor{black}{Within Mold Integration calculations, we account for three different sources of uncertainty: First, from the thermodynamic integration (TI) simulations, in which we determine $<N(\varepsilon)>$. We evaluate the standard deviation of $<N(\varepsilon)>$ by performing a block analysis in which we split the whole trajectory (typically of the order of 400 ns) into 20 blocks, so that each block can be considered independent. Then, we approximate the uncertainty of each $<N(\varepsilon)>$ data point through their typical standard deviation. Secondly, we consider the associated uncertainty due to the steep change from the intermediate/high occupation regime to the almost completely occupied regime across TI (i.e., from $\varepsilon$ values of 8.75 to 9.25 $k_B$T in Fig. 1(c) of the main text and Fig. \ref{hist} for the (11$\overline{2}$) crystal orientation). The associated error to this behaviour is determined by performing simulations starting from two different types of initial configurations. One in which all the potential wells are occupied, and other in which the mold potential wells are relatively empty. Then, by performing TI across the whole range of $\varepsilon$ from the two different initial configurations, one can establish the region affected by hysteresis. Such region is highlighted in yellow in Figure \ref{hist}, where we show the same integration as in Figure 1(c) of the main text, but adding grey symbols which correspond to simulations in which the potential wells of the initial configuration were initially unoccupied. The associated uncertainty to the observed hysteresis along the integration pathway in $\gamma_s$ corresponds to $\sim$4 mJ/m$^2$ (less than 3\% of $\gamma_s$). The sum of these two uncertainties here described gives us the error bars represented in Figure 2(a) of the main text for each estimate of $\gamma_s$ at $r_w > r_w^o$ (filled symbols). We note that for computing $\gamma_s$ we employ the values of $<N(\varepsilon)>$ obtained from the integration pathway performed with the initial configuration in which the potential wells were already filled. The reason behind that is that the kinetics of mold voiding are usually faster than those of mold filling, hence contributing for the system to reach quicker the equilibrium value of $<N(\varepsilon)>$ \cite{espinosa2014moldkk}. Finally, the third source of error in our $\gamma_s$ estimates come from the determination of the optimal well radius ($r_w^o$), and the extrapolation of $\gamma_s$ to $r_w^o$. As discussed in the main text, we need to determine the interval in which $r_w^o$ lays (Figure 1(b) of the main text), to later extrapolate $\gamma_s(r_w)$ to $r_w^o$ and ensure reversibility as much as possible across the integration pathway. This last source of uncertainty can be narrowed down by increasing the grid of $r_w$ to determine $r_w^o$, and by performing the thermodynamic integration at $r_w > r_w^o$ as close as possible to $r_w^o$, so the extrapolation of $\gamma_s$ is minimal. The combination of these three sources of error led us to a total uncertainty in the interfacial free energy of each crystal orientation of roughly 10-15 mJ/m$^2$, which approximately represents a 10\% of the interfacial free energy. In relative value, such uncertainty is similar to that found for Lennard-Jones particles \cite{espinosa2014moldkk}, NaCl with its melt \cite{espinosa2015crystalkk}, water  \cite{espinosa2016icekk} or hard-spheres \cite{sanchez2021fcckk} when evaluating $\gamma_s$ through the Mold Integration technique. }

\section{Crystal configurations to generate the different molds}

Here we provide the crystal configurations that we employ to generate the different molds in GROMOS-96 (.g96) GROMACS format, where distances are given in nm.

\subsection{(100)}

Given the following configuration, the (100) Miller index (as well as the (010) and (001) orientations) corresponds to any Cartesian direction (x, y or z)

\begin{Verbatim}[frame=single]
TITLE
water_salt
END
POSITION
    1 Na    Na         1    0.000000000    0.000000000    0.000000000
    2 Na    Na         2    0.291500000    0.291500000    0.000000000
    3 Na    Na         3    0.291500000    0.000000000    0.291500000
    4 Na    Na         4    0.000000000    0.291500000    0.291500000
    5 Cl    Cl         5    0.291500000    0.291500000    0.291500000
    6 Cl    Cl         6    0.291500000    0.000000000    0.000000000
    7 Cl    Cl         7    0.000000000    0.000000000    0.291500000
    8 Cl    Cl         8    0.000000000    0.291500000    0.000000000
END
BOX
    0.583000000    0.583000000    0.583000000
END
\end{Verbatim}

\subsection{(110)}

Given the following configuration, the (110) Miller index corresponds to the y Cartesian direction.

\begin{Verbatim}[frame=single]
TITLE
water_salt
END
POSITION
    1 Na    Na         1    0.000000000    0.000000000    0.000000000
    2 Na    Na         2    0.204410428    0.204410428    0.289080000
    3 Cl    Cl         3    0.000000000    0.000000000    0.289080000
    4 Cl    Cl         4    0.204410428    0.204410428    0.000000000
END
BOX
    0.408820856    0.408820856    0.578160000
END
\end{Verbatim}

\subsection{(111)}

Given the following configuration, the (111) Miller index corresponds to the z Cartesian direction.

\begin{Verbatim}[frame=single]
TITLE
water_salt
END
POSITION
    1 Na    Na         1    0.000000000    0.000000000    0.000000000
    2 Na    Na         2    0.204410428    0.354049246    0.000000000
    3 Cl    Cl         3    0.204410428    0.118016415    0.166897629
    4 Cl    Cl         4    0.000000000    0.472065662    0.166897629

END
BOX
    0.408820856    0.708098493    0.333795258
END
\end{Verbatim}

\subsection{(11$\overline{2}$)}

Given the following configuration, the (11$\overline{2}$) Miller index corresponds to the z Cartesian direction.

\begin{Verbatim}[frame=single]
TITLE
water_salt
END
POSITION
 1    Na    Na      1       0.000000000    0.000000000    0.000000000
 2    Cl    Cl      2       0.000000000    0.500701250    0.000000000
 3    Cl    Cl      3       0.204410430    0.833450208    0.118016412
 4    Na    Na      4       0.204410430    0.333800833    0.118016412
 5    Na    Na      5       0.000000000    0.667601666    0.236032824
 6    Cl    Cl      6       0.000000000    0.166900416    0.236032824
 7    Cl    Cl      7       0.204410430    0.500701250    0.354049236
 8    Na    Na      8       0.204410430    0.000000000    0.354049236
 9    Cl    Cl      9       0.000000000    0.834502083    0.472065648
 10   Na    Na      10      0.000000000    0.333800833    0.472065648
 11   Na    Na      11      0.204410430    0.667601666    0.590082060
 12   Cl    Cl      12      0.204410430    0.166900416    0.590082060
END
BOX
    0.408820860    1.001402500    0.708098472
END
\end{Verbatim}

\subsection{MI configurations}

\begin{figure}
    \centering
        \begin{tabular}{c c c}
            (100) & & (110) \vspace{0.2cm}\\
            \includegraphics[width=0.35\linewidth]{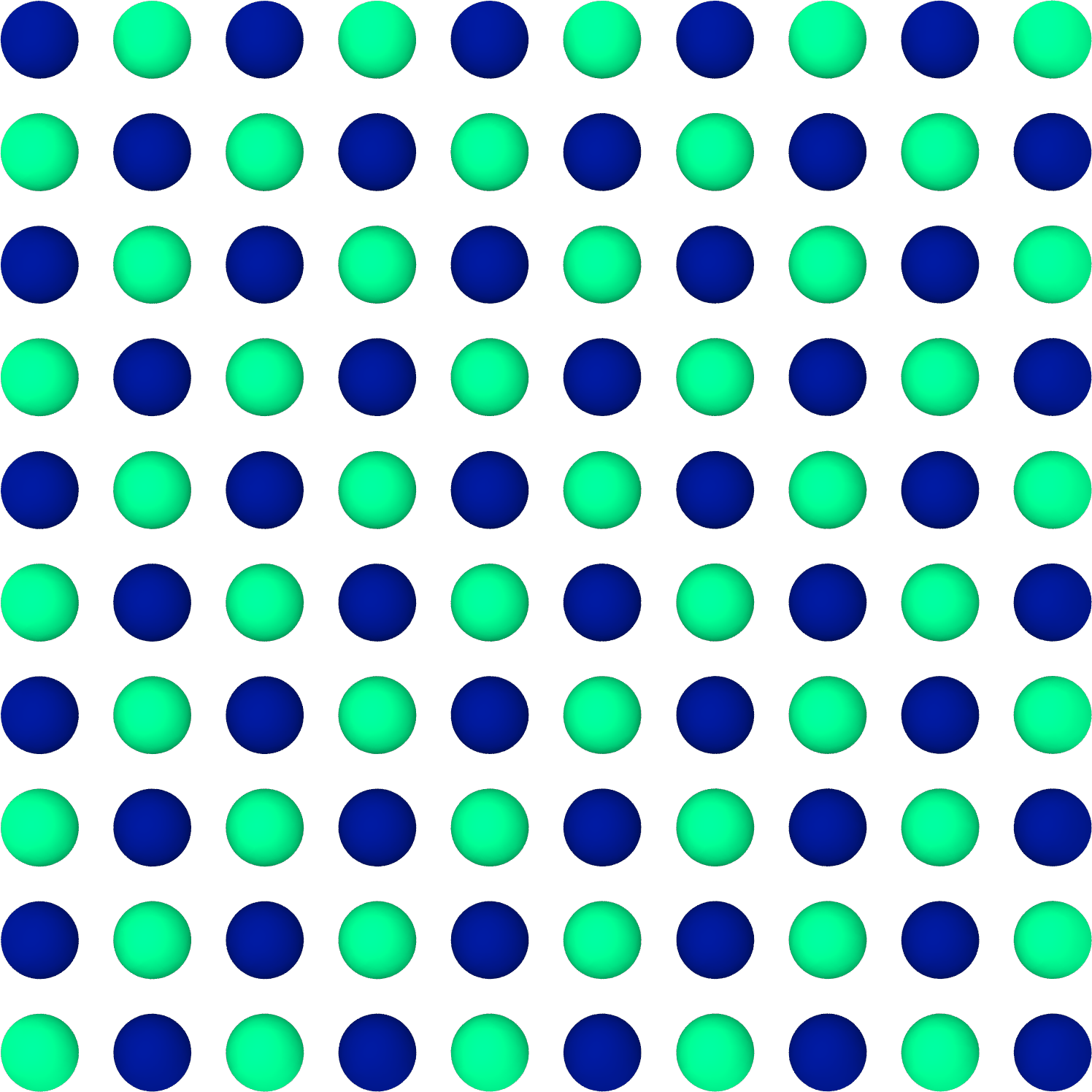} & \ & \includegraphics[width=0.35\linewidth]{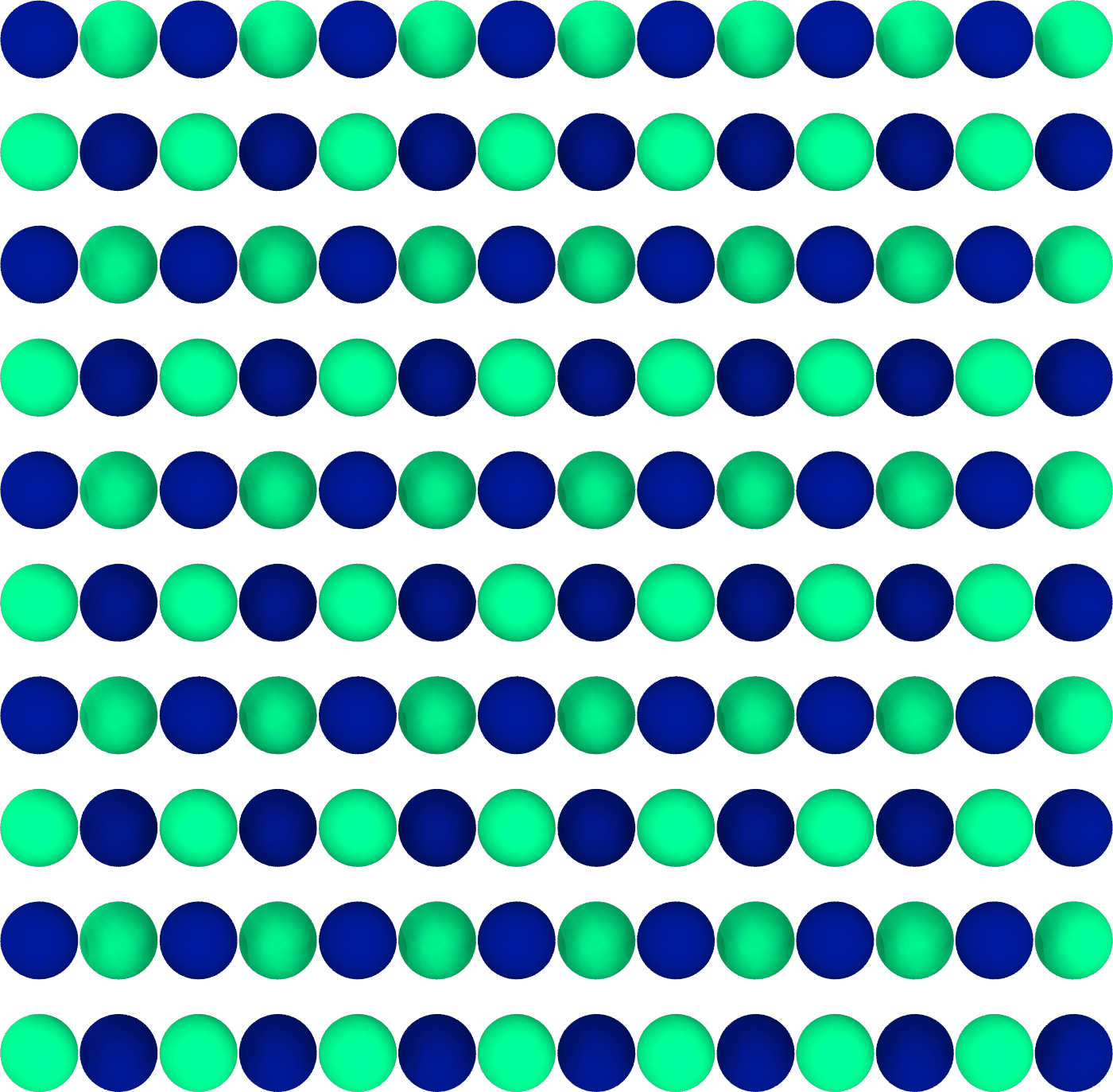} \vspace{0.2cm}\\
            (111) & & (11$\overline{2}$) \vspace{0.2cm}\\
            \includegraphics[width=0.35\linewidth]{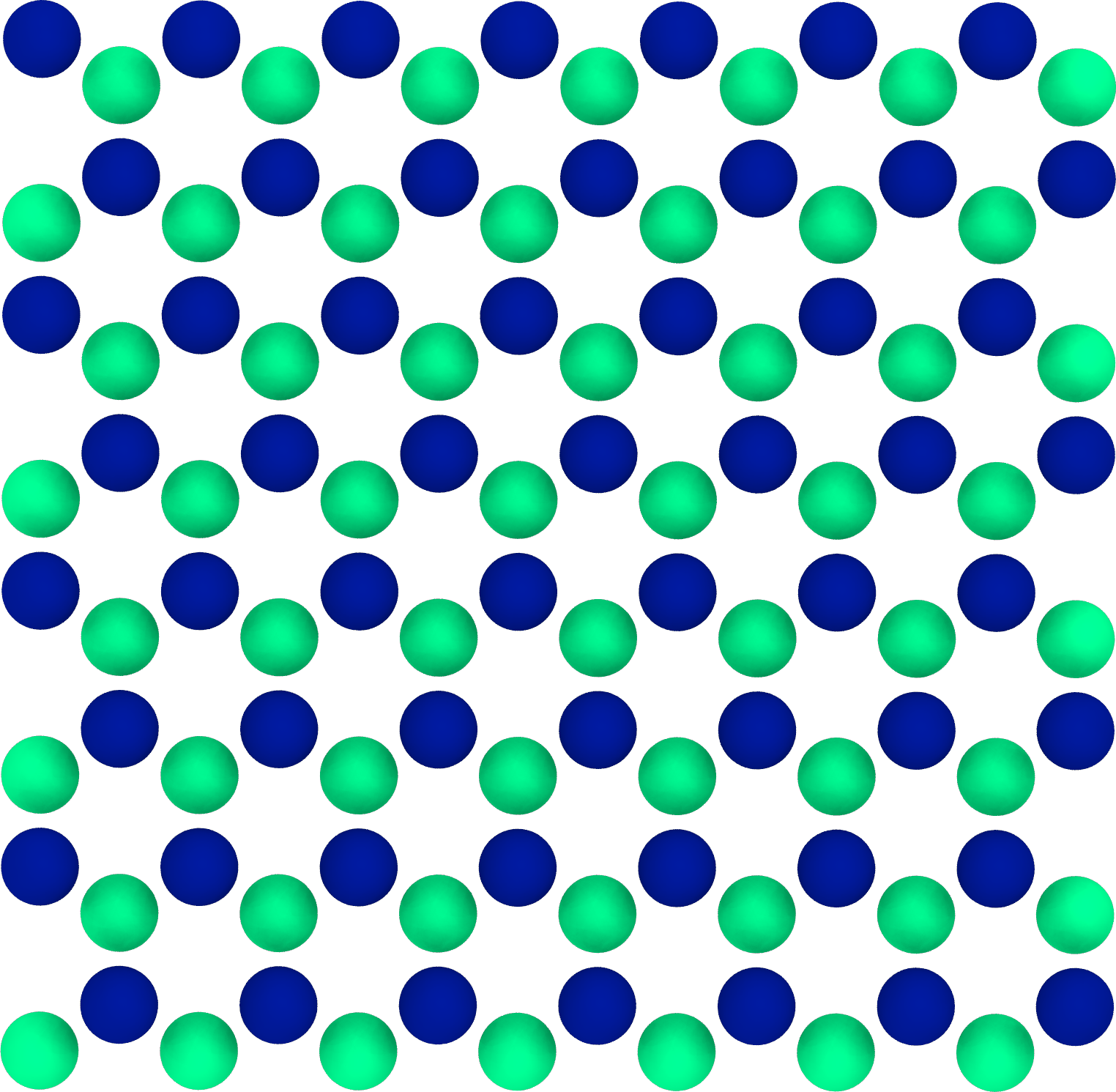} & \ & \includegraphics[width=0.35\linewidth]{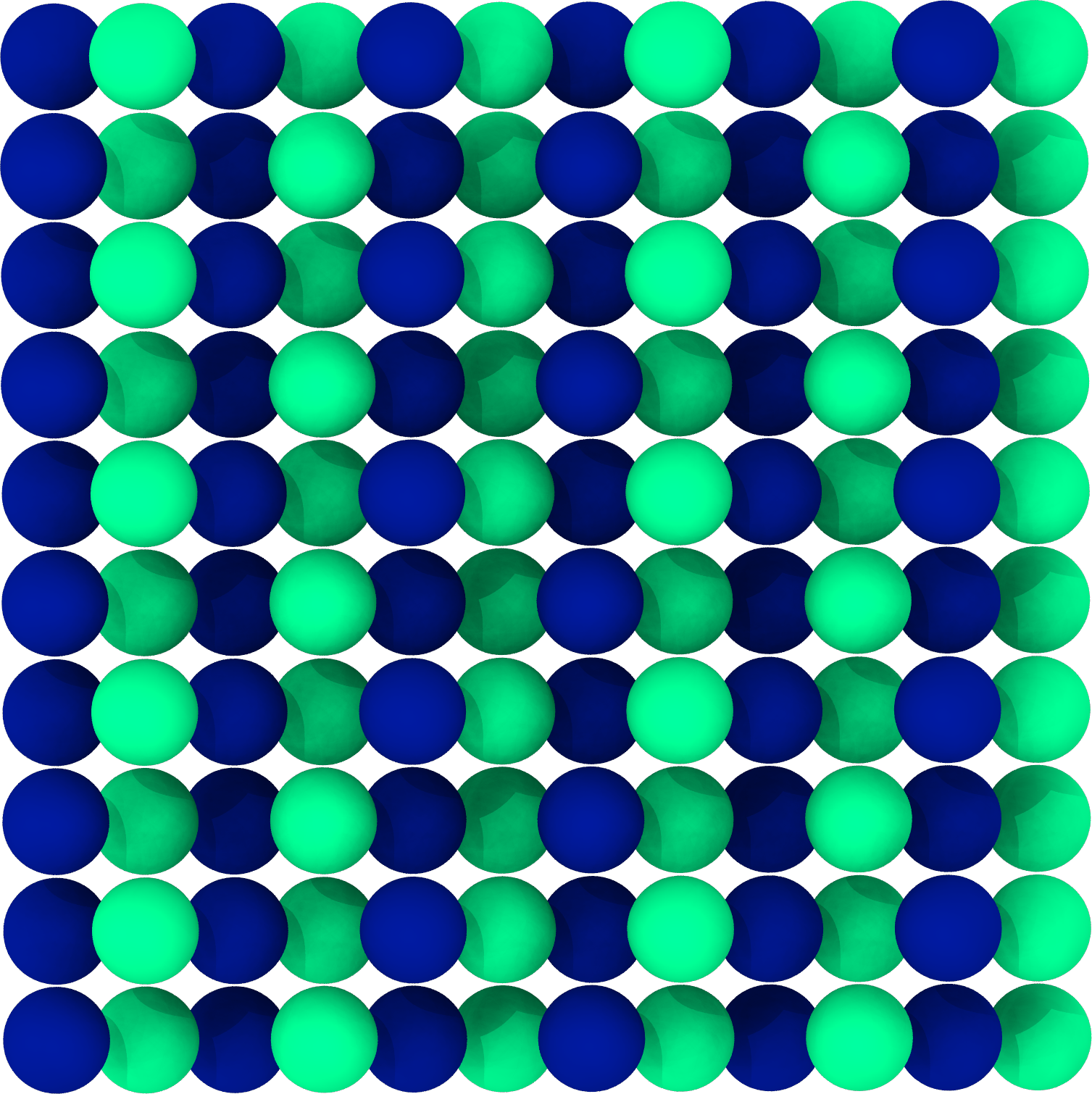}
        \end{tabular}
    \caption{\textcolor{black}{Snapshots of the different NaCl crystal orientations studied here. Please note that multiple layers of the distinct crystal faces are included. Blue spheres account for Na$^+$ ions, and green spheres for Cl$^-$ ions.}}
    \label{fotos}
\end{figure}

The full MI simulation boxes, along with the necessary files to run the simulations in GROMACS 4.6.7 have been uploaded to \href{https://github.com/ignacio-sb/NaCl_MI}{the} following Github repository: [\href{https://github.com/ignacio-sb/NaCl_MI}{https://github.com/ignacio-sb/NaCl\_MI}]. In our files, the potential wells for Na$^+$ and Cl$^-$ ions correspond to N and P atoms respectively. \textcolor{black}{We show images of the employed planes from a perpendicular point of view to the interface in Fig. \ref{fotos}.}

\section{$\bar{q}_4$-$\bar{q}_6$ local order parameter}

To determine the number of solid-like ions belonging to the crystal slab as a function of time (as shown in Fig. 1b of the main text), we make use of the $\bar{q}_6$ local order parameter proposed by Lechner and Dellago \cite{lechner2008accuratekk}. For technical details on this parameter please see Ref. \cite{lechner2008accuratekk}. In our analysis using the $\bar{q}_6$ parameter, water is not included in the evaluation of the order parameter, and cations and anions are treated as identical particles. Moreover, the nearest neighbor cut-off distance is set to 3.53 \r{A}. As a difference to Ref. \cite{lechner2008accuratekk}, and as recently proposed in Ref. \cite{lamas2021homogeneouskk} for NaCl/water calculations, we employ an unnormalized $\bar{q}_6$ parameter to distinguish between solid-like ions and ions within the solution. Making use of the mislabeling criterion \cite{lamas2021homogeneouskk,espinosa2016seedingkk,sanchez2022homogeneouskk,sanchez2021parasitickk}, we set the threshold at $\bar{q}_6=1.5$, where ions with $\bar{q}_6>1.5$ are considered to be solid-like. Nevertheless, we note that for our MI calculations, detecting an extremely accurate number of solid-like ions within the crystal slab is not critical rather than hinting whether the slab is just growing or melting \cite{espinosa2014moldkk}. 

{\color{black}
\section{Wulff's construction}}

\textcolor{black}{We make use of the Wulff construction method \cite{wulff1901xxvkk} to determine the equilibrium shape of the macroscopic crystal at the saturation concentration. Through this method, we can estimate the weight of each crystal orientation to the overall shape of the crystal. Furthermore, it can provide an average value of $\gamma_s$ for the equilibrium crystal at the saturation concentration. To perform these calculations, we employ the open source WulffPack software \cite{rahm2020wulffpackkk}. Starting from an octahedron configuration and introducing the different values of $\gamma_s$ obtained through our MI calculations as an input, we obtain a resulting crystal which is mainly formed by the (100) plane (83.7\%), with the corners of an octahedron cut out by the (11$\overline{2}$) plane (contributing the remaining 16.3\%). The resulting average $\gamma_{s,W}$ can be estimated through the following equation:} 

\begin{equation} 
\textcolor{black}{\gamma_{s,W}= \frac{\sum_i \gamma_{s,i} A_i}{\sum_i A_i}}
\end{equation}

\begin{figure}
    \centering
    \includegraphics[width=0.55\linewidth]{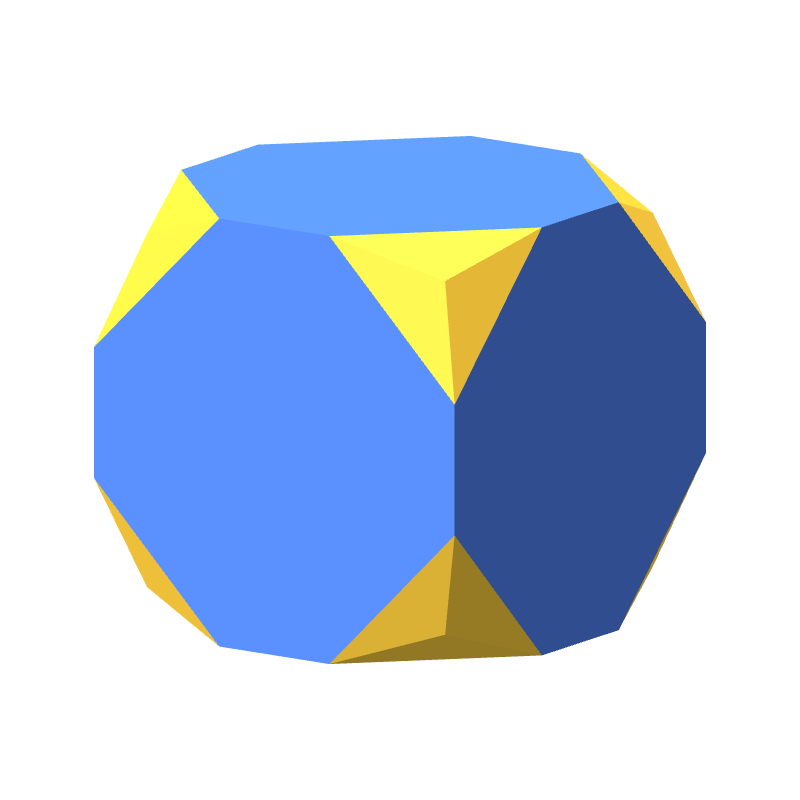}
    \vspace*{-0.5cm}
    \caption{\textcolor{black}{Predicted shape of the equilibrium crystal at the saturation concentration obtained through the Wulff's construction using the interfacial free energies computed via MI calculations for the (100), (110), (111), and (11$\overline{2}$) planes. WulffPack software \cite{rahm2020wulffpackkk} has been used to perform these calculations. The blue surface corresponds to the area in which the (100) plane is exposed, while the yellow surface depicts the area in which the (11$\overline{2}$) plane appears. Please note that knowledge of the interfacial free energy of facets with higher Miller indices may modify this result.}}
    \label{foto}
\end{figure}

\textcolor{black}{where $\gamma_{s,i}$ refers to the interfacial free energy of a given crystal orientation, and $A_i$ to the corresponding area of such plane across the macroscopic crystallite. By means of the Wulff's construction, we find a value of $\gamma_{s,W}$=109.5 mJ/m$^2$. Furthermore, a snapshot showing the obtained crystal shape through this method at the saturation concentration is presented in Figure \ref{foto}.}

\begin{thebibliography}{66}%
\makeatletter
\providecommand \@ifxundefined [1]{%
 \@ifx{#1\undefined}
}%
\providecommand \@ifnum [1]{%
 \ifnum #1\expandafter \@firstoftwo
 \else \expandafter \@secondoftwo
 \fi
}%
\providecommand \@ifx [1]{%
 \ifx #1\expandafter \@firstoftwo
 \else \expandafter \@secondoftwo
 \fi
}%
\providecommand \natexlab [1]{#1}%
\providecommand \enquote  [1]{``#1''}%
\providecommand \bibnamefont  [1]{#1}%
\providecommand \bibfnamefont [1]{#1}%
\providecommand \citenamefont [1]{#1}%
\providecommand \href@noop [0]{\@secondoftwo}%
\providecommand \href [0]{\begingroup \@sanitize@url \@href}%
\providecommand \@href[1]{\@@startlink{#1}\@@href}%
\providecommand \@@href[1]{\endgroup#1\@@endlink}%
\providecommand \@sanitize@url [0]{\catcode `\\12\catcode `\$12\catcode
  `\&12\catcode `\#12\catcode `\^12\catcode `\_12\catcode `\%12\relax}%
\providecommand \@@startlink[1]{}%
\providecommand \@@endlink[0]{}%
\providecommand \url  [0]{\begingroup\@sanitize@url \@url }%
\providecommand \@url [1]{\endgroup\@href {#1}{\urlprefix }}%
\providecommand \urlprefix  [0]{URL }%
\providecommand \Eprint [0]{\href }%
\providecommand \doibase [0]{http://dx.doi.org/}%
\providecommand \selectlanguage [0]{\@gobble}%
\providecommand \bibinfo  [0]{\@secondoftwo}%
\providecommand \bibfield  [0]{\@secondoftwo}%
\providecommand \translation [1]{[#1]}%
\providecommand \BibitemOpen [0]{}%
\providecommand \bibitemStop [0]{}%
\providecommand \bibitemNoStop [0]{.\EOS\space}%
\providecommand \EOS [0]{\spacefactor3000\relax}%
\providecommand \BibitemShut  [1]{\csname bibitem#1\endcsname}%
\let\auto@bib@innerbib\@empty
\bibitem [{\citenamefont {Lyman}\ and\ \citenamefont
  {Fleming}(1940)}]{lyman1940composition}%
  \BibitemOpen
  \bibfield  {author} {\bibinfo {author} {\bibfnamefont {J.}~\bibnamefont
  {Lyman}}\ and\ \bibinfo {author} {\bibfnamefont {R.~H.}\ \bibnamefont
  {Fleming}},\ }\href@noop {} {\bibfield  {journal} {\bibinfo  {journal} {J.
  mar. Res}\ }\textbf {\bibinfo {volume} {3}},\ \bibinfo {pages} {134}
  (\bibinfo {year} {1940})}\BibitemShut {NoStop}%
\bibitem [{\citenamefont {P{\"o}schl}(2005)}]{poschl2005atmospheric}%
  \BibitemOpen
  \bibfield  {author} {\bibinfo {author} {\bibfnamefont {U.}~\bibnamefont
  {P{\"o}schl}},\ }\href@noop {} {\bibfield  {journal} {\bibinfo  {journal}
  {Angewandte Chemie International Edition}\ }\textbf {\bibinfo {volume}
  {44}},\ \bibinfo {pages} {7520} (\bibinfo {year} {2005})}\BibitemShut
  {NoStop}%
\bibitem [{\citenamefont {Martin}(2000)}]{martin2000phase}%
  \BibitemOpen
  \bibfield  {author} {\bibinfo {author} {\bibfnamefont {S.~T.}\ \bibnamefont
  {Martin}},\ }\href@noop {} {\bibfield  {journal} {\bibinfo  {journal}
  {Chemical Reviews}\ }\textbf {\bibinfo {volume} {100}},\ \bibinfo {pages}
  {3403} (\bibinfo {year} {2000})}\BibitemShut {NoStop}%
\bibitem [{\citenamefont {Ickes}\ \emph {et~al.}(2015)\citenamefont {Ickes},
  \citenamefont {Welti}, \citenamefont {Hoose},\ and\ \citenamefont
  {Lohmann}}]{ickes2015classical}%
  \BibitemOpen
  \bibfield  {author} {\bibinfo {author} {\bibfnamefont {L.}~\bibnamefont
  {Ickes}}, \bibinfo {author} {\bibfnamefont {A.}~\bibnamefont {Welti}},
  \bibinfo {author} {\bibfnamefont {C.}~\bibnamefont {Hoose}}, \ and\ \bibinfo
  {author} {\bibfnamefont {U.}~\bibnamefont {Lohmann}},\ }\href@noop {}
  {\bibfield  {journal} {\bibinfo  {journal} {Physical Chemistry Chemical
  Physics}\ }\textbf {\bibinfo {volume} {17}},\ \bibinfo {pages} {5514}
  (\bibinfo {year} {2015})}\BibitemShut {NoStop}%
\bibitem [{\citenamefont {Bahadur}\ \emph {et~al.}(2007)\citenamefont
  {Bahadur}, \citenamefont {Russell},\ and\ \citenamefont
  {Alavi}}]{bahadur2007surface}%
  \BibitemOpen
  \bibfield  {author} {\bibinfo {author} {\bibfnamefont {R.}~\bibnamefont
  {Bahadur}}, \bibinfo {author} {\bibfnamefont {L.~M.}\ \bibnamefont
  {Russell}}, \ and\ \bibinfo {author} {\bibfnamefont {S.}~\bibnamefont
  {Alavi}},\ }\href@noop {} {\bibfield  {journal} {\bibinfo  {journal} {The
  Journal of Physical Chemistry B}\ }\textbf {\bibinfo {volume} {111}},\
  \bibinfo {pages} {11989} (\bibinfo {year} {2007})}\BibitemShut {NoStop}%
\bibitem [{\citenamefont {Broughton}\ and\ \citenamefont
  {Gilmer}(1986)}]{broughton1986molecular}%
  \BibitemOpen
  \bibfield  {author} {\bibinfo {author} {\bibfnamefont {J.~Q.}\ \bibnamefont
  {Broughton}}\ and\ \bibinfo {author} {\bibfnamefont {G.~H.}\ \bibnamefont
  {Gilmer}},\ }\href@noop {} {\bibfield  {journal} {\bibinfo  {journal} {The
  Journal of chemical physics}\ }\textbf {\bibinfo {volume} {84}},\ \bibinfo
  {pages} {5759} (\bibinfo {year} {1986})}\BibitemShut {NoStop}%
\bibitem [{\citenamefont {Fern{\'a}ndez}\ \emph {et~al.}(2012)\citenamefont
  {Fern{\'a}ndez}, \citenamefont {Martin-Mayor}, \citenamefont {Seoane},\ and\
  \citenamefont {Verrocchio}}]{fernandez2012equilibrium}%
  \BibitemOpen
  \bibfield  {author} {\bibinfo {author} {\bibfnamefont {L.}~\bibnamefont
  {Fern{\'a}ndez}}, \bibinfo {author} {\bibfnamefont {V.}~\bibnamefont
  {Martin-Mayor}}, \bibinfo {author} {\bibfnamefont {B.}~\bibnamefont
  {Seoane}}, \ and\ \bibinfo {author} {\bibfnamefont {P.}~\bibnamefont
  {Verrocchio}},\ }\href@noop {} {\bibfield  {journal} {\bibinfo  {journal}
  {Physical review letters}\ }\textbf {\bibinfo {volume} {108}},\ \bibinfo
  {pages} {165701} (\bibinfo {year} {2012})}\BibitemShut {NoStop}%
\bibitem [{\citenamefont {Angioletti-Uberti}\ \emph {et~al.}(2010)\citenamefont
  {Angioletti-Uberti}, \citenamefont {Ceriotti}, \citenamefont {Lee},\ and\
  \citenamefont {Finnis}}]{angioletti2010solid}%
  \BibitemOpen
  \bibfield  {author} {\bibinfo {author} {\bibfnamefont {S.}~\bibnamefont
  {Angioletti-Uberti}}, \bibinfo {author} {\bibfnamefont {M.}~\bibnamefont
  {Ceriotti}}, \bibinfo {author} {\bibfnamefont {P.~D.}\ \bibnamefont {Lee}}, \
  and\ \bibinfo {author} {\bibfnamefont {M.~W.}\ \bibnamefont {Finnis}},\
  }\href@noop {} {\bibfield  {journal} {\bibinfo  {journal} {Physical Review
  B}\ }\textbf {\bibinfo {volume} {81}},\ \bibinfo {pages} {125416} (\bibinfo
  {year} {2010})}\BibitemShut {NoStop}%
\bibitem [{\citenamefont {Espinosa}\ \emph {et~al.}(2014)\citenamefont
  {Espinosa}, \citenamefont {Vega},\ and\ \citenamefont
  {Sanz}}]{espinosa2014mold}%
  \BibitemOpen
  \bibfield  {author} {\bibinfo {author} {\bibfnamefont {J.}~\bibnamefont
  {Espinosa}}, \bibinfo {author} {\bibfnamefont {C.}~\bibnamefont {Vega}}, \
  and\ \bibinfo {author} {\bibfnamefont {E.}~\bibnamefont {Sanz}},\ }\href@noop
  {} {\bibfield  {journal} {\bibinfo  {journal} {The Journal of chemical
  physics}\ }\textbf {\bibinfo {volume} {141}},\ \bibinfo {pages} {134709}
  (\bibinfo {year} {2014})}\BibitemShut {NoStop}%
\bibitem [{\citenamefont {Hoyt}\ \emph {et~al.}(2001)\citenamefont {Hoyt},
  \citenamefont {Asta},\ and\ \citenamefont {Karma}}]{hoyt2001method}%
  \BibitemOpen
  \bibfield  {author} {\bibinfo {author} {\bibfnamefont {J.}~\bibnamefont
  {Hoyt}}, \bibinfo {author} {\bibfnamefont {M.}~\bibnamefont {Asta}}, \ and\
  \bibinfo {author} {\bibfnamefont {A.}~\bibnamefont {Karma}},\ }\href@noop {}
  {\bibfield  {journal} {\bibinfo  {journal} {Physical review letters}\
  }\textbf {\bibinfo {volume} {86}},\ \bibinfo {pages} {5530} (\bibinfo {year}
  {2001})}\BibitemShut {NoStop}%
\bibitem [{\citenamefont {Benjamin}\ and\ \citenamefont
  {Horbach}(2014)}]{benjamin2014crystal}%
  \BibitemOpen
  \bibfield  {author} {\bibinfo {author} {\bibfnamefont {R.}~\bibnamefont
  {Benjamin}}\ and\ \bibinfo {author} {\bibfnamefont {J.}~\bibnamefont
  {Horbach}},\ }\href@noop {} {\bibfield  {journal} {\bibinfo  {journal} {The
  Journal of chemical physics}\ }\textbf {\bibinfo {volume} {141}},\ \bibinfo
  {pages} {044715} (\bibinfo {year} {2014})}\BibitemShut {NoStop}%
\bibitem [{\citenamefont {Schilling}\ and\ \citenamefont
  {Schmid}(2009)}]{schilling2009computing}%
  \BibitemOpen
  \bibfield  {author} {\bibinfo {author} {\bibfnamefont {T.}~\bibnamefont
  {Schilling}}\ and\ \bibinfo {author} {\bibfnamefont {F.}~\bibnamefont
  {Schmid}},\ }\href@noop {} {\bibfield  {journal} {\bibinfo  {journal} {The
  Journal of chemical physics}\ }\textbf {\bibinfo {volume} {131}},\ \bibinfo
  {pages} {231102} (\bibinfo {year} {2009})}\BibitemShut {NoStop}%
\bibitem [{\citenamefont {B{\"u}ltmann}\ and\ \citenamefont
  {Schilling}(2020)}]{bultmann2020computation}%
  \BibitemOpen
  \bibfield  {author} {\bibinfo {author} {\bibfnamefont {M.}~\bibnamefont
  {B{\"u}ltmann}}\ and\ \bibinfo {author} {\bibfnamefont {T.}~\bibnamefont
  {Schilling}},\ }\href@noop {} {\bibfield  {journal} {\bibinfo  {journal}
  {Physical Review E}\ }\textbf {\bibinfo {volume} {102}},\ \bibinfo {pages}
  {042123} (\bibinfo {year} {2020})}\BibitemShut {NoStop}%
\bibitem [{\citenamefont {Sanchez-Burgos}\ \emph
  {et~al.}(2021{\natexlab{a}})\citenamefont {Sanchez-Burgos}, \citenamefont
  {Sanz}, \citenamefont {Vega},\ and\ \citenamefont
  {Espinosa}}]{sanchez2021fcc}%
  \BibitemOpen
  \bibfield  {author} {\bibinfo {author} {\bibfnamefont {I.}~\bibnamefont
  {Sanchez-Burgos}}, \bibinfo {author} {\bibfnamefont {E.}~\bibnamefont
  {Sanz}}, \bibinfo {author} {\bibfnamefont {C.}~\bibnamefont {Vega}}, \ and\
  \bibinfo {author} {\bibfnamefont {J.~R.}\ \bibnamefont {Espinosa}},\
  }\href@noop {} {\bibfield  {journal} {\bibinfo  {journal} {Physical Chemistry
  Chemical Physics}\ }\textbf {\bibinfo {volume} {23}},\ \bibinfo {pages}
  {19611} (\bibinfo {year} {2021}{\natexlab{a}})}\BibitemShut {NoStop}%
\bibitem [{\citenamefont {Espinosa}\ \emph {et~al.}(2015)\citenamefont
  {Espinosa}, \citenamefont {Vega}, \citenamefont {Valeriani},\ and\
  \citenamefont {Sanz}}]{espinosa2015crystal}%
  \BibitemOpen
  \bibfield  {author} {\bibinfo {author} {\bibfnamefont {J.~R.}\ \bibnamefont
  {Espinosa}}, \bibinfo {author} {\bibfnamefont {C.}~\bibnamefont {Vega}},
  \bibinfo {author} {\bibfnamefont {C.}~\bibnamefont {Valeriani}}, \ and\
  \bibinfo {author} {\bibfnamefont {E.}~\bibnamefont {Sanz}},\ }\href@noop {}
  {\bibfield  {journal} {\bibinfo  {journal} {The Journal of Chemical Physics}\
  }\textbf {\bibinfo {volume} {142}},\ \bibinfo {pages} {194709} (\bibinfo
  {year} {2015})}\BibitemShut {NoStop}%
\bibitem [{\citenamefont {Espinosa}\ \emph
  {et~al.}(2016{\natexlab{a}})\citenamefont {Espinosa}, \citenamefont {Vega},\
  and\ \citenamefont {Sanz}}]{espinosa2016ice}%
  \BibitemOpen
  \bibfield  {author} {\bibinfo {author} {\bibfnamefont {J.~R.}\ \bibnamefont
  {Espinosa}}, \bibinfo {author} {\bibfnamefont {C.}~\bibnamefont {Vega}}, \
  and\ \bibinfo {author} {\bibfnamefont {E.}~\bibnamefont {Sanz}},\ }\href@noop
  {} {\bibfield  {journal} {\bibinfo  {journal} {The Journal of Physical
  Chemistry C}\ }\textbf {\bibinfo {volume} {120}},\ \bibinfo {pages} {8068}
  (\bibinfo {year} {2016}{\natexlab{a}})}\BibitemShut {NoStop}%
\bibitem [{\citenamefont {Soria}\ \emph {et~al.}(2018)\citenamefont {Soria},
  \citenamefont {Espinosa}, \citenamefont {Ramirez}, \citenamefont {Valeriani},
  \citenamefont {Vega},\ and\ \citenamefont {Sanz}}]{soria2018simulation}%
  \BibitemOpen
  \bibfield  {author} {\bibinfo {author} {\bibfnamefont {G.~D.}\ \bibnamefont
  {Soria}}, \bibinfo {author} {\bibfnamefont {J.~R.}\ \bibnamefont {Espinosa}},
  \bibinfo {author} {\bibfnamefont {J.}~\bibnamefont {Ramirez}}, \bibinfo
  {author} {\bibfnamefont {C.}~\bibnamefont {Valeriani}}, \bibinfo {author}
  {\bibfnamefont {C.}~\bibnamefont {Vega}}, \ and\ \bibinfo {author}
  {\bibfnamefont {E.}~\bibnamefont {Sanz}},\ }\href@noop {} {\bibfield
  {journal} {\bibinfo  {journal} {The Journal of chemical physics}\ }\textbf
  {\bibinfo {volume} {148}},\ \bibinfo {pages} {222811} (\bibinfo {year}
  {2018})}\BibitemShut {NoStop}%
\bibitem [{\citenamefont {Benet}\ \emph {et~al.}(2015)\citenamefont {Benet},
  \citenamefont {MacDowell},\ and\ \citenamefont
  {Sanz}}]{benet2015interfacial}%
  \BibitemOpen
  \bibfield  {author} {\bibinfo {author} {\bibfnamefont {J.}~\bibnamefont
  {Benet}}, \bibinfo {author} {\bibfnamefont {L.~G.}\ \bibnamefont
  {MacDowell}}, \ and\ \bibinfo {author} {\bibfnamefont {E.}~\bibnamefont
  {Sanz}},\ }\href@noop {} {\bibfield  {journal} {\bibinfo  {journal} {The
  Journal of Chemical Physics}\ }\textbf {\bibinfo {volume} {142}},\ \bibinfo
  {pages} {134706} (\bibinfo {year} {2015})}\BibitemShut {NoStop}%
\bibitem [{\citenamefont {Davidchack}\ and\ \citenamefont
  {Laird}(2003)}]{davidchack2003direct}%
  \BibitemOpen
  \bibfield  {author} {\bibinfo {author} {\bibfnamefont {R.~L.}\ \bibnamefont
  {Davidchack}}\ and\ \bibinfo {author} {\bibfnamefont {B.~B.}\ \bibnamefont
  {Laird}},\ }\href@noop {} {\bibfield  {journal} {\bibinfo  {journal} {The
  Journal of chemical physics}\ }\textbf {\bibinfo {volume} {118}},\ \bibinfo
  {pages} {7651} (\bibinfo {year} {2003})}\BibitemShut {NoStop}%
\bibitem [{\citenamefont {Davidchack}(2010)}]{davidchack2010hard}%
  \BibitemOpen
  \bibfield  {author} {\bibinfo {author} {\bibfnamefont {R.~L.}\ \bibnamefont
  {Davidchack}},\ }\href@noop {} {\bibfield  {journal} {\bibinfo  {journal}
  {The Journal of chemical physics}\ }\textbf {\bibinfo {volume} {133}},\
  \bibinfo {pages} {234701} (\bibinfo {year} {2010})}\BibitemShut {NoStop}%
\bibitem [{\citenamefont {Ambler}\ \emph {et~al.}(2017)\citenamefont {Ambler},
  \citenamefont {Vorselaars}, \citenamefont {Allen},\ and\ \citenamefont
  {Quigley}}]{ambler2017solid}%
  \BibitemOpen
  \bibfield  {author} {\bibinfo {author} {\bibfnamefont {M.}~\bibnamefont
  {Ambler}}, \bibinfo {author} {\bibfnamefont {B.}~\bibnamefont {Vorselaars}},
  \bibinfo {author} {\bibfnamefont {M.~P.}\ \bibnamefont {Allen}}, \ and\
  \bibinfo {author} {\bibfnamefont {D.}~\bibnamefont {Quigley}},\ }\href@noop
  {} {\bibfield  {journal} {\bibinfo  {journal} {The Journal of Chemical
  Physics}\ }\textbf {\bibinfo {volume} {146}},\ \bibinfo {pages} {074701}
  (\bibinfo {year} {2017})}\BibitemShut {NoStop}%
\bibitem [{\citenamefont {Asta}\ \emph {et~al.}(2002)\citenamefont {Asta},
  \citenamefont {Hoyt},\ and\ \citenamefont {Karma}}]{asta2002calculation}%
  \BibitemOpen
  \bibfield  {author} {\bibinfo {author} {\bibfnamefont {M.}~\bibnamefont
  {Asta}}, \bibinfo {author} {\bibfnamefont {J.}~\bibnamefont {Hoyt}}, \ and\
  \bibinfo {author} {\bibfnamefont {A.}~\bibnamefont {Karma}},\ }\href@noop {}
  {\bibfield  {journal} {\bibinfo  {journal} {Physical Review B}\ }\textbf
  {\bibinfo {volume} {66}},\ \bibinfo {pages} {100101} (\bibinfo {year}
  {2002})}\BibitemShut {NoStop}%
\bibitem [{\citenamefont {Algaba}\ \emph {et~al.}(2022)\citenamefont {Algaba},
  \citenamefont {Acu{\~n}a}, \citenamefont {M{\'\i}guez}, \citenamefont
  {Mendiboure}, \citenamefont {Zer{\'o}n},\ and\ \citenamefont
  {Blas}}]{algaba2022simulation}%
  \BibitemOpen
  \bibfield  {author} {\bibinfo {author} {\bibfnamefont {J.}~\bibnamefont
  {Algaba}}, \bibinfo {author} {\bibfnamefont {E.}~\bibnamefont {Acu{\~n}a}},
  \bibinfo {author} {\bibfnamefont {J.~M.}\ \bibnamefont {M{\'\i}guez}},
  \bibinfo {author} {\bibfnamefont {B.}~\bibnamefont {Mendiboure}}, \bibinfo
  {author} {\bibfnamefont {I.~M.}\ \bibnamefont {Zer{\'o}n}}, \ and\ \bibinfo
  {author} {\bibfnamefont {F.~J.}\ \bibnamefont {Blas}},\ }\href@noop {}
  {\bibfield  {journal} {\bibinfo  {journal} {Journal of Colloid and Interface
  Science}\ } (\bibinfo {year} {2022})}\BibitemShut {NoStop}%
\bibitem [{\citenamefont {Lamas}\ \emph {et~al.}(2021)\citenamefont {Lamas},
  \citenamefont {Espinosa}, \citenamefont {Conde}, \citenamefont
  {Ram{\'\i}rez}, \citenamefont {de~Hijes}, \citenamefont {Noya}, \citenamefont
  {Vega},\ and\ \citenamefont {Sanz}}]{lamas2021homogeneous}%
  \BibitemOpen
  \bibfield  {author} {\bibinfo {author} {\bibfnamefont {C.}~\bibnamefont
  {Lamas}}, \bibinfo {author} {\bibfnamefont {J.}~\bibnamefont {Espinosa}},
  \bibinfo {author} {\bibfnamefont {M.}~\bibnamefont {Conde}}, \bibinfo
  {author} {\bibfnamefont {J.}~\bibnamefont {Ram{\'\i}rez}}, \bibinfo {author}
  {\bibfnamefont {P.~M.}\ \bibnamefont {de~Hijes}}, \bibinfo {author}
  {\bibfnamefont {E.~G.}\ \bibnamefont {Noya}}, \bibinfo {author}
  {\bibfnamefont {C.}~\bibnamefont {Vega}}, \ and\ \bibinfo {author}
  {\bibfnamefont {E.}~\bibnamefont {Sanz}},\ }\href@noop {} {\bibfield
  {journal} {\bibinfo  {journal} {Physical Chemistry Chemical Physics}\
  }\textbf {\bibinfo {volume} {23}},\ \bibinfo {pages} {26843} (\bibinfo {year}
  {2021})}\BibitemShut {NoStop}%
\bibitem [{\citenamefont {Jiang}\ \emph {et~al.}(2018)\citenamefont {Jiang},
  \citenamefont {Haji-Akbari}, \citenamefont {Debenedetti},\ and\ \citenamefont
  {Panagiotopoulos}}]{jiang2018forward}%
  \BibitemOpen
  \bibfield  {author} {\bibinfo {author} {\bibfnamefont {H.}~\bibnamefont
  {Jiang}}, \bibinfo {author} {\bibfnamefont {A.}~\bibnamefont {Haji-Akbari}},
  \bibinfo {author} {\bibfnamefont {P.~G.}\ \bibnamefont {Debenedetti}}, \ and\
  \bibinfo {author} {\bibfnamefont {A.~Z.}\ \bibnamefont {Panagiotopoulos}},\
  }\href@noop {} {\bibfield  {journal} {\bibinfo  {journal} {The Journal of
  chemical physics}\ }\textbf {\bibinfo {volume} {148}},\ \bibinfo {pages}
  {044505} (\bibinfo {year} {2018})}\BibitemShut {NoStop}%
\bibitem [{\citenamefont {Na}\ \emph {et~al.}(1994)\citenamefont {Na},
  \citenamefont {Arnold},\ and\ \citenamefont {Myerson}}]{na1994cluster}%
  \BibitemOpen
  \bibfield  {author} {\bibinfo {author} {\bibfnamefont {H.-S.}\ \bibnamefont
  {Na}}, \bibinfo {author} {\bibfnamefont {S.}~\bibnamefont {Arnold}}, \ and\
  \bibinfo {author} {\bibfnamefont {A.~S.}\ \bibnamefont {Myerson}},\
  }\href@noop {} {\bibfield  {journal} {\bibinfo  {journal} {Journal of crystal
  growth}\ }\textbf {\bibinfo {volume} {139}},\ \bibinfo {pages} {104}
  (\bibinfo {year} {1994})}\BibitemShut {NoStop}%
\bibitem [{\citenamefont {Volmer}\ and\ \citenamefont
  {Weber}(1926)}]{ZPC_1926_119_277_nolotengo}%
  \BibitemOpen
  \bibfield  {author} {\bibinfo {author} {\bibfnamefont {M.}~\bibnamefont
  {Volmer}}\ and\ \bibinfo {author} {\bibfnamefont {A.}~\bibnamefont {Weber}},\
  }\href@noop {} {\bibfield  {journal} {\bibinfo  {journal} {Z. Phys. Chem.}\
  }\textbf {\bibinfo {volume} {119}},\ \bibinfo {pages} {277} (\bibinfo {year}
  {1926})}\BibitemShut {NoStop}%
\bibitem [{\citenamefont {Becker}\ and\ \citenamefont
  {Doring}(1935)}]{becker-doring}%
  \BibitemOpen
  \bibfield  {author} {\bibinfo {author} {\bibfnamefont {R.}~\bibnamefont
  {Becker}}\ and\ \bibinfo {author} {\bibfnamefont {W.}~\bibnamefont
  {Doring}},\ }\href@noop {} {\bibfield  {journal} {\bibinfo  {journal} {Ann.
  Phys.}\ }\textbf {\bibinfo {volume} {416}},\ \bibinfo {pages} {719} (\bibinfo
  {year} {1935})}\BibitemShut {NoStop}%
\bibitem [{\citenamefont {Zimmermann}\ \emph {et~al.}(2018)\citenamefont
  {Zimmermann}, \citenamefont {Vorselaars}, \citenamefont {Espinosa},
  \citenamefont {Quigley}, \citenamefont {Smith}, \citenamefont {Sanz},
  \citenamefont {Vega},\ and\ \citenamefont {Peters}}]{zimmermann2018nacl}%
  \BibitemOpen
  \bibfield  {author} {\bibinfo {author} {\bibfnamefont {N.~E.}\ \bibnamefont
  {Zimmermann}}, \bibinfo {author} {\bibfnamefont {B.}~\bibnamefont
  {Vorselaars}}, \bibinfo {author} {\bibfnamefont {J.~R.}\ \bibnamefont
  {Espinosa}}, \bibinfo {author} {\bibfnamefont {D.}~\bibnamefont {Quigley}},
  \bibinfo {author} {\bibfnamefont {W.~R.}\ \bibnamefont {Smith}}, \bibinfo
  {author} {\bibfnamefont {E.}~\bibnamefont {Sanz}}, \bibinfo {author}
  {\bibfnamefont {C.}~\bibnamefont {Vega}}, \ and\ \bibinfo {author}
  {\bibfnamefont {B.}~\bibnamefont {Peters}},\ }\href@noop {} {\bibfield
  {journal} {\bibinfo  {journal} {The Journal of chemical physics}\ }\textbf
  {\bibinfo {volume} {148}},\ \bibinfo {pages} {222838} (\bibinfo {year}
  {2018})}\BibitemShut {NoStop}%
\bibitem [{\citenamefont {Berendsen}\ \emph {et~al.}(1987)\citenamefont
  {Berendsen}, \citenamefont {Grigera},\ and\ \citenamefont
  {Straatsma}}]{berendsen1987missing}%
  \BibitemOpen
  \bibfield  {author} {\bibinfo {author} {\bibfnamefont {H.}~\bibnamefont
  {Berendsen}}, \bibinfo {author} {\bibfnamefont {J.}~\bibnamefont {Grigera}},
  \ and\ \bibinfo {author} {\bibfnamefont {T.}~\bibnamefont {Straatsma}},\
  }\href@noop {} {\bibfield  {journal} {\bibinfo  {journal} {Journal of
  Physical Chemistry}\ }\textbf {\bibinfo {volume} {91}},\ \bibinfo {pages}
  {6269} (\bibinfo {year} {1987})}\BibitemShut {NoStop}%
\bibitem [{\citenamefont {Joung}\ and\ \citenamefont
  {Cheatham~III}(2009)}]{joung2009molecular}%
  \BibitemOpen
  \bibfield  {author} {\bibinfo {author} {\bibfnamefont {I.~S.}\ \bibnamefont
  {Joung}}\ and\ \bibinfo {author} {\bibfnamefont {T.~E.}\ \bibnamefont
  {Cheatham~III}},\ }\href@noop {} {\bibfield  {journal} {\bibinfo  {journal}
  {The Journal of Physical Chemistry B}\ }\textbf {\bibinfo {volume} {113}},\
  \bibinfo {pages} {13279} (\bibinfo {year} {2009})}\BibitemShut {NoStop}%
\bibitem [{\citenamefont {Lorentz}(1881)}]{lorentz1881ueber}%
  \BibitemOpen
  \bibfield  {author} {\bibinfo {author} {\bibfnamefont {H.~A.}\ \bibnamefont
  {Lorentz}},\ }\href@noop {} {\bibfield  {journal} {\bibinfo  {journal}
  {Annalen der physik}\ }\textbf {\bibinfo {volume} {248}},\ \bibinfo {pages}
  {127} (\bibinfo {year} {1881})}\BibitemShut {NoStop}%
\bibitem [{\citenamefont {Berthelot}(1899)}]{berthelot1899methode}%
  \BibitemOpen
  \bibfield  {author} {\bibinfo {author} {\bibfnamefont {D.}~\bibnamefont
  {Berthelot}},\ }\href@noop {} {\bibfield  {journal} {\bibinfo  {journal}
  {Journal de Physique Th{\'e}orique et Appliqu{\'e}e}\ }\textbf {\bibinfo
  {volume} {8}},\ \bibinfo {pages} {263} (\bibinfo {year} {1899})}\BibitemShut
  {NoStop}%
\bibitem [{\citenamefont {Zeron}\ \emph {et~al.}(2022)\citenamefont {Zeron},
  \citenamefont {M{\'\i}guez}, \citenamefont {Mendiboure}, \citenamefont
  {Algaba},\ and\ \citenamefont {Blas}}]{zeron2022simulation}%
  \BibitemOpen
  \bibfield  {author} {\bibinfo {author} {\bibfnamefont {I.~M.}\ \bibnamefont
  {Zeron}}, \bibinfo {author} {\bibfnamefont {J.~M.}\ \bibnamefont
  {M{\'\i}guez}}, \bibinfo {author} {\bibfnamefont {B.}~\bibnamefont
  {Mendiboure}}, \bibinfo {author} {\bibfnamefont {J.}~\bibnamefont {Algaba}},
  \ and\ \bibinfo {author} {\bibfnamefont {F.~J.}\ \bibnamefont {Blas}},\
  }\href@noop {} {\bibfield  {journal} {\bibinfo  {journal} {The Journal of
  Chemical Physics}\ } (\bibinfo {year} {2022})}\BibitemShut {NoStop}%
\bibitem [{\citenamefont {Bekker}\ \emph {et~al.}(1993)\citenamefont {Bekker},
  \citenamefont {Berendsen}, \citenamefont {Dijkstra}, \citenamefont
  {Achterop}, \citenamefont {Vondrumen}, \citenamefont {VANDERSPOEL},
  \citenamefont {Sijbers}, \citenamefont {Keegstra},\ and\ \citenamefont
  {Renardus}}]{bekker1993gromacs}%
  \BibitemOpen
  \bibfield  {author} {\bibinfo {author} {\bibfnamefont {H.}~\bibnamefont
  {Bekker}}, \bibinfo {author} {\bibfnamefont {H.}~\bibnamefont {Berendsen}},
  \bibinfo {author} {\bibfnamefont {E.}~\bibnamefont {Dijkstra}}, \bibinfo
  {author} {\bibfnamefont {S.}~\bibnamefont {Achterop}}, \bibinfo {author}
  {\bibfnamefont {R.}~\bibnamefont {Vondrumen}}, \bibinfo {author}
  {\bibfnamefont {D.}~\bibnamefont {VANDERSPOEL}}, \bibinfo {author}
  {\bibfnamefont {A.}~\bibnamefont {Sijbers}}, \bibinfo {author} {\bibfnamefont
  {H.}~\bibnamefont {Keegstra}}, \ and\ \bibinfo {author} {\bibfnamefont
  {M.}~\bibnamefont {Renardus}},\ }in\ \href@noop {} {\emph {\bibinfo
  {booktitle} {4th International Conference on Computational Physics (PC
  92)}}}\ (\bibinfo {organization} {World Scientific Publishing},\ \bibinfo
  {year} {1993})\ pp.\ \bibinfo {pages} {252--256}\BibitemShut {NoStop}%
\bibitem [{\citenamefont {Bussi}\ \emph {et~al.}(2007)\citenamefont {Bussi},
  \citenamefont {Donadio},\ and\ \citenamefont
  {Parrinello}}]{bussi2007canonical}%
  \BibitemOpen
  \bibfield  {author} {\bibinfo {author} {\bibfnamefont {G.}~\bibnamefont
  {Bussi}}, \bibinfo {author} {\bibfnamefont {D.}~\bibnamefont {Donadio}}, \
  and\ \bibinfo {author} {\bibfnamefont {M.}~\bibnamefont {Parrinello}},\
  }\href@noop {} {\bibfield  {journal} {\bibinfo  {journal} {The Journal of
  chemical physics}\ }\textbf {\bibinfo {volume} {126}},\ \bibinfo {pages}
  {014101} (\bibinfo {year} {2007})}\BibitemShut {NoStop}%
\bibitem [{\citenamefont {Parrinello}\ and\ \citenamefont
  {Rahman}(1981)}]{parrinello1981polymorphic}%
  \BibitemOpen
  \bibfield  {author} {\bibinfo {author} {\bibfnamefont {M.}~\bibnamefont
  {Parrinello}}\ and\ \bibinfo {author} {\bibfnamefont {A.}~\bibnamefont
  {Rahman}},\ }\href@noop {} {\bibfield  {journal} {\bibinfo  {journal}
  {Journal of Applied physics}\ }\textbf {\bibinfo {volume} {52}},\ \bibinfo
  {pages} {7182} (\bibinfo {year} {1981})}\BibitemShut {NoStop}%
\bibitem [{\citenamefont {Hockney}\ \emph {et~al.}(1974)\citenamefont
  {Hockney}, \citenamefont {Goel},\ and\ \citenamefont
  {Eastwood}}]{hockney1974quiet}%
  \BibitemOpen
  \bibfield  {author} {\bibinfo {author} {\bibfnamefont {R.~W.}\ \bibnamefont
  {Hockney}}, \bibinfo {author} {\bibfnamefont {S.}~\bibnamefont {Goel}}, \
  and\ \bibinfo {author} {\bibfnamefont {J.}~\bibnamefont {Eastwood}},\
  }\href@noop {} {\bibfield  {journal} {\bibinfo  {journal} {Journal of
  Computational Physics}\ }\textbf {\bibinfo {volume} {14}},\ \bibinfo {pages}
  {148} (\bibinfo {year} {1974})}\BibitemShut {NoStop}%
\bibitem [{\citenamefont {Darden}\ \emph {et~al.}(1993)\citenamefont {Darden},
  \citenamefont {York},\ and\ \citenamefont {Pedersen}}]{darden1993particle}%
  \BibitemOpen
  \bibfield  {author} {\bibinfo {author} {\bibfnamefont {T.}~\bibnamefont
  {Darden}}, \bibinfo {author} {\bibfnamefont {D.}~\bibnamefont {York}}, \ and\
  \bibinfo {author} {\bibfnamefont {L.}~\bibnamefont {Pedersen}},\ }\href@noop
  {} {\bibfield  {journal} {\bibinfo  {journal} {The Journal of chemical
  physics}\ }\textbf {\bibinfo {volume} {98}},\ \bibinfo {pages} {10089}
  (\bibinfo {year} {1993})}\BibitemShut {NoStop}%
\bibitem [{\citenamefont {Essmann}\ \emph {et~al.}(1995)\citenamefont
  {Essmann}, \citenamefont {Perera}, \citenamefont {Berkowitz}, \citenamefont
  {Darden}, \citenamefont {Lee},\ and\ \citenamefont
  {Pedersen}}]{essmann1995smooth}%
  \BibitemOpen
  \bibfield  {author} {\bibinfo {author} {\bibfnamefont {U.}~\bibnamefont
  {Essmann}}, \bibinfo {author} {\bibfnamefont {L.}~\bibnamefont {Perera}},
  \bibinfo {author} {\bibfnamefont {M.~L.}\ \bibnamefont {Berkowitz}}, \bibinfo
  {author} {\bibfnamefont {T.}~\bibnamefont {Darden}}, \bibinfo {author}
  {\bibfnamefont {H.}~\bibnamefont {Lee}}, \ and\ \bibinfo {author}
  {\bibfnamefont {L.~G.}\ \bibnamefont {Pedersen}},\ }\href@noop {} {\bibfield
  {journal} {\bibinfo  {journal} {The Journal of chemical physics}\ }\textbf
  {\bibinfo {volume} {103}},\ \bibinfo {pages} {8577} (\bibinfo {year}
  {1995})}\BibitemShut {NoStop}%
\bibitem [{\citenamefont {Hess}\ \emph {et~al.}(1997)\citenamefont {Hess},
  \citenamefont {Bekker}, \citenamefont {Berendsen},\ and\ \citenamefont
  {Fraaije}}]{hess1997lincs}%
  \BibitemOpen
  \bibfield  {author} {\bibinfo {author} {\bibfnamefont {B.}~\bibnamefont
  {Hess}}, \bibinfo {author} {\bibfnamefont {H.}~\bibnamefont {Bekker}},
  \bibinfo {author} {\bibfnamefont {H.~J.}\ \bibnamefont {Berendsen}}, \ and\
  \bibinfo {author} {\bibfnamefont {J.~G.}\ \bibnamefont {Fraaije}},\
  }\href@noop {} {\bibfield  {journal} {\bibinfo  {journal} {Journal of
  computational chemistry}\ }\textbf {\bibinfo {volume} {18}},\ \bibinfo
  {pages} {1463} (\bibinfo {year} {1997})}\BibitemShut {NoStop}%
\bibitem [{\citenamefont {Espinosa}\ \emph
  {et~al.}(2016{\natexlab{b}})\citenamefont {Espinosa}, \citenamefont {Vega},
  \citenamefont {Valeriani},\ and\ \citenamefont {Sanz}}]{espinosa2016seeding}%
  \BibitemOpen
  \bibfield  {author} {\bibinfo {author} {\bibfnamefont {J.~R.}\ \bibnamefont
  {Espinosa}}, \bibinfo {author} {\bibfnamefont {C.}~\bibnamefont {Vega}},
  \bibinfo {author} {\bibfnamefont {C.}~\bibnamefont {Valeriani}}, \ and\
  \bibinfo {author} {\bibfnamefont {E.}~\bibnamefont {Sanz}},\ }\href@noop {}
  {\bibfield  {journal} {\bibinfo  {journal} {The Journal of chemical physics}\
  }\textbf {\bibinfo {volume} {144}},\ \bibinfo {pages} {034501} (\bibinfo
  {year} {2016}{\natexlab{b}})}\BibitemShut {NoStop}%
\bibitem [{\citenamefont {Sanchez-Burgos}\ \emph {et~al.}(2022)\citenamefont
  {Sanchez-Burgos}, \citenamefont {Tejedor}, \citenamefont {Vega},
  \citenamefont {Conde}, \citenamefont {Sanz}, \citenamefont {Ramirez},\ and\
  \citenamefont {Espinosa}}]{sanchez2022homogeneous}%
  \BibitemOpen
  \bibfield  {author} {\bibinfo {author} {\bibfnamefont {I.}~\bibnamefont
  {Sanchez-Burgos}}, \bibinfo {author} {\bibfnamefont {A.~R.}\ \bibnamefont
  {Tejedor}}, \bibinfo {author} {\bibfnamefont {C.}~\bibnamefont {Vega}},
  \bibinfo {author} {\bibfnamefont {M.~M.}\ \bibnamefont {Conde}}, \bibinfo
  {author} {\bibfnamefont {E.}~\bibnamefont {Sanz}}, \bibinfo {author}
  {\bibfnamefont {J.}~\bibnamefont {Ramirez}}, \ and\ \bibinfo {author}
  {\bibfnamefont {J.~R.}\ \bibnamefont {Espinosa}},\ }\href@noop {} {\bibfield
  {journal} {\bibinfo  {journal} {The Journal of Chemical Physics}\ }\textbf
  {\bibinfo {volume} {157}},\ \bibinfo {pages} {094503} (\bibinfo {year}
  {2022})}\BibitemShut {NoStop}%
\bibitem [{\citenamefont {Sanchez-Burgos}\ \emph
  {et~al.}(2021{\natexlab{b}})\citenamefont {Sanchez-Burgos}, \citenamefont
  {Garaizar}, \citenamefont {Vega}, \citenamefont {Sanz},\ and\ \citenamefont
  {Espinosa}}]{sanchez2021parasitic}%
  \BibitemOpen
  \bibfield  {author} {\bibinfo {author} {\bibfnamefont {I.}~\bibnamefont
  {Sanchez-Burgos}}, \bibinfo {author} {\bibfnamefont {A.}~\bibnamefont
  {Garaizar}}, \bibinfo {author} {\bibfnamefont {C.}~\bibnamefont {Vega}},
  \bibinfo {author} {\bibfnamefont {E.}~\bibnamefont {Sanz}}, \ and\ \bibinfo
  {author} {\bibfnamefont {J.~R.}\ \bibnamefont {Espinosa}},\ }\href@noop {}
  {\bibfield  {journal} {\bibinfo  {journal} {Soft Matter}\ }\textbf {\bibinfo
  {volume} {17}},\ \bibinfo {pages} {489} (\bibinfo {year}
  {2021}{\natexlab{b}})}\BibitemShut {NoStop}%
\bibitem [{\citenamefont {Joung}\ and\ \citenamefont
  {Cheatham~III}(2008)}]{joung2008determination}%
  \BibitemOpen
  \bibfield  {author} {\bibinfo {author} {\bibfnamefont {I.~S.}\ \bibnamefont
  {Joung}}\ and\ \bibinfo {author} {\bibfnamefont {T.~E.}\ \bibnamefont
  {Cheatham~III}},\ }\href@noop {} {\bibfield  {journal} {\bibinfo  {journal}
  {The journal of physical chemistry B}\ }\textbf {\bibinfo {volume} {112}},\
  \bibinfo {pages} {9020} (\bibinfo {year} {2008})}\BibitemShut {NoStop}%
\bibitem [{\citenamefont {Orozco}\ \emph {et~al.}(2014)\citenamefont {Orozco},
  \citenamefont {Moultos}, \citenamefont {Jiang}, \citenamefont {Economou},\
  and\ \citenamefont {Panagiotopoulos}}]{orozco2014molecular}%
  \BibitemOpen
  \bibfield  {author} {\bibinfo {author} {\bibfnamefont {G.~A.}\ \bibnamefont
  {Orozco}}, \bibinfo {author} {\bibfnamefont {O.~A.}\ \bibnamefont {Moultos}},
  \bibinfo {author} {\bibfnamefont {H.}~\bibnamefont {Jiang}}, \bibinfo
  {author} {\bibfnamefont {I.~G.}\ \bibnamefont {Economou}}, \ and\ \bibinfo
  {author} {\bibfnamefont {A.~Z.}\ \bibnamefont {Panagiotopoulos}},\
  }\href@noop {} {\bibfield  {journal} {\bibinfo  {journal} {The Journal of
  chemical physics}\ }\textbf {\bibinfo {volume} {141}},\ \bibinfo {pages}
  {234507} (\bibinfo {year} {2014})}\BibitemShut {NoStop}%
\bibitem [{\citenamefont {Benavides}\ \emph {et~al.}(2016)\citenamefont
  {Benavides}, \citenamefont {Aragones},\ and\ \citenamefont
  {Vega}}]{benavides2016consensus}%
  \BibitemOpen
  \bibfield  {author} {\bibinfo {author} {\bibfnamefont {A.}~\bibnamefont
  {Benavides}}, \bibinfo {author} {\bibfnamefont {J.}~\bibnamefont {Aragones}},
  \ and\ \bibinfo {author} {\bibfnamefont {C.}~\bibnamefont {Vega}},\
  }\href@noop {} {\bibfield  {journal} {\bibinfo  {journal} {The Journal of
  Chemical Physics}\ }\textbf {\bibinfo {volume} {144}},\ \bibinfo {pages}
  {124504} (\bibinfo {year} {2016})}\BibitemShut {NoStop}%
\bibitem [{\citenamefont {Espinosa}\ \emph
  {et~al.}(2016{\natexlab{c}})\citenamefont {Espinosa}, \citenamefont {Young},
  \citenamefont {Jiang}, \citenamefont {Gupta}, \citenamefont {Vega},
  \citenamefont {Sanz}, \citenamefont {Debenedetti},\ and\ \citenamefont
  {Panagiotopoulos}}]{espinosa2016calculation}%
  \BibitemOpen
  \bibfield  {author} {\bibinfo {author} {\bibfnamefont {J.}~\bibnamefont
  {Espinosa}}, \bibinfo {author} {\bibfnamefont {J.}~\bibnamefont {Young}},
  \bibinfo {author} {\bibfnamefont {H.}~\bibnamefont {Jiang}}, \bibinfo
  {author} {\bibfnamefont {D.}~\bibnamefont {Gupta}}, \bibinfo {author}
  {\bibfnamefont {C.}~\bibnamefont {Vega}}, \bibinfo {author} {\bibfnamefont
  {E.}~\bibnamefont {Sanz}}, \bibinfo {author} {\bibfnamefont {P.~G.}\
  \bibnamefont {Debenedetti}}, \ and\ \bibinfo {author} {\bibfnamefont {A.~Z.}\
  \bibnamefont {Panagiotopoulos}},\ }\href@noop {} {\bibfield  {journal}
  {\bibinfo  {journal} {The Journal of chemical physics}\ }\textbf {\bibinfo
  {volume} {145}},\ \bibinfo {pages} {154111} (\bibinfo {year}
  {2016}{\natexlab{c}})}\BibitemShut {NoStop}%
\bibitem [{\citenamefont {Mester}\ and\ \citenamefont
  {Panagiotopoulos}(2015)}]{mester2015temperature}%
  \BibitemOpen
  \bibfield  {author} {\bibinfo {author} {\bibfnamefont {Z.}~\bibnamefont
  {Mester}}\ and\ \bibinfo {author} {\bibfnamefont {A.~Z.}\ \bibnamefont
  {Panagiotopoulos}},\ }\href@noop {} {\bibfield  {journal} {\bibinfo
  {journal} {The Journal of chemical physics}\ }\textbf {\bibinfo {volume}
  {143}},\ \bibinfo {pages} {044505} (\bibinfo {year} {2015})}\BibitemShut
  {NoStop}%
\bibitem [{\citenamefont {Mou{\v{c}}ka}\ \emph {et~al.}(2013)\citenamefont
  {Mou{\v{c}}ka}, \citenamefont {Nezbeda},\ and\ \citenamefont
  {Smith}}]{mouvcka2013molecular}%
  \BibitemOpen
  \bibfield  {author} {\bibinfo {author} {\bibfnamefont {F.}~\bibnamefont
  {Mou{\v{c}}ka}}, \bibinfo {author} {\bibfnamefont {I.}~\bibnamefont
  {Nezbeda}}, \ and\ \bibinfo {author} {\bibfnamefont {W.~R.}\ \bibnamefont
  {Smith}},\ }\href@noop {} {\bibfield  {journal} {\bibinfo  {journal} {The
  Journal of chemical physics}\ }\textbf {\bibinfo {volume} {138}},\ \bibinfo
  {pages} {154102} (\bibinfo {year} {2013})}\BibitemShut {NoStop}%
\bibitem [{\citenamefont {Nezbeda}\ \emph {et~al.}(2016)\citenamefont
  {Nezbeda}, \citenamefont {Mou{\v{c}}ka},\ and\ \citenamefont
  {Smith}}]{nezbeda2016recent}%
  \BibitemOpen
  \bibfield  {author} {\bibinfo {author} {\bibfnamefont {I.}~\bibnamefont
  {Nezbeda}}, \bibinfo {author} {\bibfnamefont {F.}~\bibnamefont
  {Mou{\v{c}}ka}}, \ and\ \bibinfo {author} {\bibfnamefont {W.~R.}\
  \bibnamefont {Smith}},\ }\href@noop {} {\bibfield  {journal} {\bibinfo
  {journal} {Molecular Physics}\ }\textbf {\bibinfo {volume} {114}},\ \bibinfo
  {pages} {1665} (\bibinfo {year} {2016})}\BibitemShut {NoStop}%
\bibitem [{\citenamefont {Stukowski}(2009)}]{stukowski2009visualization}%
  \BibitemOpen
  \bibfield  {author} {\bibinfo {author} {\bibfnamefont {A.}~\bibnamefont
  {Stukowski}},\ }\href@noop {} {\bibfield  {journal} {\bibinfo  {journal}
  {Modelling and simulation in materials science and engineering}\ }\textbf
  {\bibinfo {volume} {18}},\ \bibinfo {pages} {015012} (\bibinfo {year}
  {2009})}\BibitemShut {NoStop}%
\bibitem [{\citenamefont {Lechner}\ and\ \citenamefont
  {Dellago}(2008)}]{lechner2008accurate}%
  \BibitemOpen
  \bibfield  {author} {\bibinfo {author} {\bibfnamefont {W.}~\bibnamefont
  {Lechner}}\ and\ \bibinfo {author} {\bibfnamefont {C.}~\bibnamefont
  {Dellago}},\ }\href@noop {} {\bibfield  {journal} {\bibinfo  {journal} {The
  Journal of chemical physics}\ }\textbf {\bibinfo {volume} {129}},\ \bibinfo
  {pages} {114707} (\bibinfo {year} {2008})}\BibitemShut {NoStop}%
\bibitem [{\citenamefont {Kolafa}(2016)}]{kolafa2016solubility}%
  \BibitemOpen
  \bibfield  {author} {\bibinfo {author} {\bibfnamefont {J.}~\bibnamefont
  {Kolafa}},\ }\href@noop {} {\bibfield  {journal} {\bibinfo  {journal} {The
  Journal of chemical physics}\ }\textbf {\bibinfo {volume} {145}},\ \bibinfo
  {pages} {204509} (\bibinfo {year} {2016})}\BibitemShut {NoStop}%
\bibitem [{\citenamefont {Perego}\ \emph {et~al.}(2015)\citenamefont {Perego},
  \citenamefont {Salvalaglio},\ and\ \citenamefont
  {Parrinello}}]{perego2015molecular}%
  \BibitemOpen
  \bibfield  {author} {\bibinfo {author} {\bibfnamefont {C.}~\bibnamefont
  {Perego}}, \bibinfo {author} {\bibfnamefont {M.}~\bibnamefont {Salvalaglio}},
  \ and\ \bibinfo {author} {\bibfnamefont {M.}~\bibnamefont {Parrinello}},\
  }\href@noop {} {\bibfield  {journal} {\bibinfo  {journal} {The Journal of
  chemical physics}\ }\textbf {\bibinfo {volume} {142}},\ \bibinfo {pages}
  {144113} (\bibinfo {year} {2015})}\BibitemShut {NoStop}%
\bibitem [{\citenamefont {Karmakar}\ \emph {et~al.}(2018)\citenamefont
  {Karmakar}, \citenamefont {Piaggi}, \citenamefont {Perego},\ and\
  \citenamefont {Parrinello}}]{karmakar2018cannibalistic}%
  \BibitemOpen
  \bibfield  {author} {\bibinfo {author} {\bibfnamefont {T.}~\bibnamefont
  {Karmakar}}, \bibinfo {author} {\bibfnamefont {P.~M.}\ \bibnamefont
  {Piaggi}}, \bibinfo {author} {\bibfnamefont {C.}~\bibnamefont {Perego}}, \
  and\ \bibinfo {author} {\bibfnamefont {M.}~\bibnamefont {Parrinello}},\
  }\href@noop {} {\bibfield  {journal} {\bibinfo  {journal} {Journal of
  chemical theory and computation}\ }\textbf {\bibinfo {volume} {14}},\
  \bibinfo {pages} {2678} (\bibinfo {year} {2018})}\BibitemShut {NoStop}%
\bibitem [{\citenamefont {Zimmermann}\ \emph {et~al.}(2015)\citenamefont
  {Zimmermann}, \citenamefont {Vorselaars}, \citenamefont {Quigley},\ and\
  \citenamefont {Peters}}]{zimmermann2015nucleation}%
  \BibitemOpen
  \bibfield  {author} {\bibinfo {author} {\bibfnamefont {N.~E.}\ \bibnamefont
  {Zimmermann}}, \bibinfo {author} {\bibfnamefont {B.}~\bibnamefont
  {Vorselaars}}, \bibinfo {author} {\bibfnamefont {D.}~\bibnamefont {Quigley}},
  \ and\ \bibinfo {author} {\bibfnamefont {B.}~\bibnamefont {Peters}},\
  }\href@noop {} {\bibfield  {journal} {\bibinfo  {journal} {Journal of the
  American Chemical Society}\ }\textbf {\bibinfo {volume} {137}},\ \bibinfo
  {pages} {13352} (\bibinfo {year} {2015})}\BibitemShut {NoStop}%
\bibitem [{\citenamefont {Wulff}(1901)}]{wulff1901xxv}%
  \BibitemOpen
  \bibfield  {author} {\bibinfo {author} {\bibfnamefont {G.}~\bibnamefont
  {Wulff}},\ }\href@noop {} {\bibfield  {journal} {\bibinfo  {journal}
  {Zeitschrift f{\"u}r Kristallographie-Crystalline Materials}\ }\textbf
  {\bibinfo {volume} {34}},\ \bibinfo {pages} {449} (\bibinfo {year}
  {1901})}\BibitemShut {NoStop}%
\bibitem [{\citenamefont {Rahm}\ and\ \citenamefont
  {Erhart}(2020)}]{rahm2020wulffpack}%
  \BibitemOpen
  \bibfield  {author} {\bibinfo {author} {\bibfnamefont {J.~M.}\ \bibnamefont
  {Rahm}}\ and\ \bibinfo {author} {\bibfnamefont {P.}~\bibnamefont {Erhart}},\
  }\href@noop {} {\bibfield  {journal} {\bibinfo  {journal} {Journal of Open
  Source Software}\ }\textbf {\bibinfo {volume} {5}},\ \bibinfo {pages} {1944}
  (\bibinfo {year} {2020})}\BibitemShut {NoStop}%
\bibitem [{\citenamefont {Aquilano}\ \emph {et~al.}(2009)\citenamefont
  {Aquilano}, \citenamefont {Pastero}, \citenamefont {Bruno},\ and\
  \citenamefont {Rubbo}}]{aquilano20091}%
  \BibitemOpen
  \bibfield  {author} {\bibinfo {author} {\bibfnamefont {D.}~\bibnamefont
  {Aquilano}}, \bibinfo {author} {\bibfnamefont {L.}~\bibnamefont {Pastero}},
  \bibinfo {author} {\bibfnamefont {M.}~\bibnamefont {Bruno}}, \ and\ \bibinfo
  {author} {\bibfnamefont {M.}~\bibnamefont {Rubbo}},\ }\href@noop {}
  {\bibfield  {journal} {\bibinfo  {journal} {Journal of crystal growth}\
  }\textbf {\bibinfo {volume} {311}},\ \bibinfo {pages} {399} (\bibinfo {year}
  {2009})}\BibitemShut {NoStop}%
\bibitem [{\citenamefont {Mu}\ \emph {et~al.}(2005)\citenamefont {Mu},
  \citenamefont {Houk},\ and\ \citenamefont {Song}}]{mu2005anisotropic}%
  \BibitemOpen
  \bibfield  {author} {\bibinfo {author} {\bibfnamefont {Y.}~\bibnamefont
  {Mu}}, \bibinfo {author} {\bibfnamefont {A.}~\bibnamefont {Houk}}, \ and\
  \bibinfo {author} {\bibfnamefont {X.}~\bibnamefont {Song}},\ }\href@noop {}
  {\bibfield  {journal} {\bibinfo  {journal} {The Journal of Physical Chemistry
  B}\ }\textbf {\bibinfo {volume} {109}},\ \bibinfo {pages} {6500} (\bibinfo
  {year} {2005})}\BibitemShut {NoStop}%
\bibitem [{\citenamefont {Benjamin}\ and\ \citenamefont
  {Horbach}(2015)}]{benjamin2015crystal}%
  \BibitemOpen
  \bibfield  {author} {\bibinfo {author} {\bibfnamefont {R.}~\bibnamefont
  {Benjamin}}\ and\ \bibinfo {author} {\bibfnamefont {J.}~\bibnamefont
  {Horbach}},\ }\href@noop {} {\bibfield  {journal} {\bibinfo  {journal}
  {Physical Review E}\ }\textbf {\bibinfo {volume} {91}},\ \bibinfo {pages}
  {032410} (\bibinfo {year} {2015})}\BibitemShut {NoStop}%
\bibitem [{\citenamefont {Schmitz}\ and\ \citenamefont
  {Virnau}(2015)}]{schmitz2015ensemble}%
  \BibitemOpen
  \bibfield  {author} {\bibinfo {author} {\bibfnamefont {F.}~\bibnamefont
  {Schmitz}}\ and\ \bibinfo {author} {\bibfnamefont {P.}~\bibnamefont
  {Virnau}},\ }\href@noop {} {\bibfield  {journal} {\bibinfo  {journal} {The
  Journal of Chemical Physics}\ }\textbf {\bibinfo {volume} {142}},\ \bibinfo
  {pages} {144108} (\bibinfo {year} {2015})}\BibitemShut {NoStop}%
\bibitem [{\citenamefont {Laird}\ \emph {et~al.}(2009)\citenamefont {Laird},
  \citenamefont {Davidchack}, \citenamefont {Yang},\ and\ \citenamefont
  {Asta}}]{laird2009determination}%
  \BibitemOpen
  \bibfield  {author} {\bibinfo {author} {\bibfnamefont {B.~B.}\ \bibnamefont
  {Laird}}, \bibinfo {author} {\bibfnamefont {R.~L.}\ \bibnamefont
  {Davidchack}}, \bibinfo {author} {\bibfnamefont {Y.}~\bibnamefont {Yang}}, \
  and\ \bibinfo {author} {\bibfnamefont {M.}~\bibnamefont {Asta}},\ }\href@noop
  {} {\bibfield  {journal} {\bibinfo  {journal} {The Journal of chemical
  physics}\ }\textbf {\bibinfo {volume} {131}},\ \bibinfo {pages} {114110}
  (\bibinfo {year} {2009})}\BibitemShut {NoStop}%
\bibitem [{\citenamefont {Montero~de Hijes}\ \emph {et~al.}(2020)\citenamefont
  {Montero~de Hijes}, \citenamefont {Espinosa}, \citenamefont {Bianco},
  \citenamefont {Sanz},\ and\ \citenamefont {Vega}}]{montero2020interfacial}%
  \BibitemOpen
  \bibfield  {author} {\bibinfo {author} {\bibfnamefont {P.}~\bibnamefont
  {Montero~de Hijes}}, \bibinfo {author} {\bibfnamefont {J.~R.}\ \bibnamefont
  {Espinosa}}, \bibinfo {author} {\bibfnamefont {V.}~\bibnamefont {Bianco}},
  \bibinfo {author} {\bibfnamefont {E.}~\bibnamefont {Sanz}}, \ and\ \bibinfo
  {author} {\bibfnamefont {C.}~\bibnamefont {Vega}},\ }\href@noop {} {\bibfield
   {journal} {\bibinfo  {journal} {The Journal of Physical Chemistry C}\
  }\textbf {\bibinfo {volume} {124}},\ \bibinfo {pages} {8795} (\bibinfo {year}
  {2020})}\BibitemShut {NoStop}%
\bibitem [{\citenamefont {Quilaqueo}\ and\ \citenamefont
  {Aguilera}(2016)}]{quilaqueo2016crystallization}%
  \BibitemOpen
  \bibfield  {author} {\bibinfo {author} {\bibfnamefont {M.}~\bibnamefont
  {Quilaqueo}}\ and\ \bibinfo {author} {\bibfnamefont {J.~M.}\ \bibnamefont
  {Aguilera}},\ }\href@noop {} {\bibfield  {journal} {\bibinfo  {journal} {Food
  Research International}\ }\textbf {\bibinfo {volume} {84}},\ \bibinfo {pages}
  {143} (\bibinfo {year} {2016})}\BibitemShut {NoStop}%
\end{thebibliography}

\begin{thebibliography}{26}%
\makeatletter
\providecommand \@ifxundefined [1]{%
 \@ifx{#1\undefined}
}%
\providecommand \@ifnum [1]{%
 \ifnum #1\expandafter \@firstoftwo
 \else \expandafter \@secondoftwo
 \fi
}%
\providecommand \@ifx [1]{%
 \ifx #1\expandafter \@firstoftwo
 \else \expandafter \@secondoftwo
 \fi
}%
\providecommand \natexlab [1]{#1}%
\providecommand \enquote  [1]{``#1''}%
\providecommand \bibnamefont  [1]{#1}%
\providecommand \bibfnamefont [1]{#1}%
\providecommand \citenamefont [1]{#1}%
\providecommand \href@noop [0]{\@secondoftwo}%
\providecommand \href [0]{\begingroup \@sanitize@url \@href}%
\providecommand \@href[1]{\@@startlink{#1}\@@href}%
\providecommand \@@href[1]{\endgroup#1\@@endlink}%
\providecommand \@sanitize@url [0]{\catcode `\\12\catcode `\$12\catcode
  `\&12\catcode `\#12\catcode `\^12\catcode `\_12\catcode `\%12\relax}%
\providecommand \@@startlink[1]{}%
\providecommand \@@endlink[0]{}%
\providecommand \url  [0]{\begingroup\@sanitize@url \@url }%
\providecommand \@url [1]{\endgroup\@href {#1}{\urlprefix }}%
\providecommand \urlprefix  [0]{URL }%
\providecommand \Eprint [0]{\href }%
\providecommand \doibase [0]{http://dx.doi.org/}%
\providecommand \selectlanguage [0]{\@gobble}%
\providecommand \bibinfo  [0]{\@secondoftwo}%
\providecommand \bibfield  [0]{\@secondoftwo}%
\providecommand \translation [1]{[#1]}%
\providecommand \BibitemOpen [0]{}%
\providecommand \bibitemStop [0]{}%
\providecommand \bibitemNoStop [0]{.\EOS\space}%
\providecommand \EOS [0]{\spacefactor3000\relax}%
\providecommand \BibitemShut  [1]{\csname bibitem#1\endcsname}%
\let\auto@bib@innerbib\@empty
\bibitem [{\citenamefont {Berendsen}\ \emph {et~al.}(1987)\citenamefont
  {Berendsen}, \citenamefont {Grigera},\ and\ \citenamefont
  {Straatsma}}]{berendsen1987missingkk}%
  \BibitemOpen
  \bibfield  {author} {\bibinfo {author} {\bibfnamefont {H.}~\bibnamefont
  {Berendsen}}, \bibinfo {author} {\bibfnamefont {J.}~\bibnamefont {Grigera}},
  \ and\ \bibinfo {author} {\bibfnamefont {T.}~\bibnamefont {Straatsma}},\
  }\href@noop {} {\bibfield  {journal} {\bibinfo  {journal} {Journal of
  Physical Chemistry}\ }\textbf {\bibinfo {volume} {91}},\ \bibinfo {pages}
  {6269} (\bibinfo {year} {1987})}\BibitemShut {NoStop}%
\bibitem [{\citenamefont {Joung}\ and\ \citenamefont
  {Cheatham~III}(2009)}]{joung2009molecularkk}%
  \BibitemOpen
  \bibfield  {author} {\bibinfo {author} {\bibfnamefont {I.~S.}\ \bibnamefont
  {Joung}}\ and\ \bibinfo {author} {\bibfnamefont {T.~E.}\ \bibnamefont
  {Cheatham~III}},\ }\href@noop {} {\bibfield  {journal} {\bibinfo  {journal}
  {The Journal of Physical Chemistry B}\ }\textbf {\bibinfo {volume} {113}},\
  \bibinfo {pages} {13279} (\bibinfo {year} {2009})}\BibitemShut {NoStop}%
\bibitem [{\citenamefont {Lorentz}(1881)}]{lorentz1881ueberkk}%
  \BibitemOpen
  \bibfield  {author} {\bibinfo {author} {\bibfnamefont {H.~A.}\ \bibnamefont
  {Lorentz}},\ }\href@noop {} {\bibfield  {journal} {\bibinfo  {journal}
  {Annalen der physik}\ }\textbf {\bibinfo {volume} {248}},\ \bibinfo {pages}
  {127} (\bibinfo {year} {1881})}\BibitemShut {NoStop}%
\bibitem [{\citenamefont {Berthelot}(1899)}]{berthelot1899methodekk}%
  \BibitemOpen
  \bibfield  {author} {\bibinfo {author} {\bibfnamefont {D.}~\bibnamefont
  {Berthelot}},\ }\href@noop {} {\bibfield  {journal} {\bibinfo  {journal}
  {Journal de Physique Th{\'e}orique et Appliqu{\'e}e}\ }\textbf {\bibinfo
  {volume} {8}},\ \bibinfo {pages} {263} (\bibinfo {year} {1899})}\BibitemShut
  {NoStop}%
\bibitem [{\citenamefont {Espinosa}\ \emph {et~al.}(2014)\citenamefont
  {Espinosa}, \citenamefont {Vega},\ and\ \citenamefont
  {Sanz}}]{espinosa2014moldkk}%
  \BibitemOpen
  \bibfield  {author} {\bibinfo {author} {\bibfnamefont {J.}~\bibnamefont
  {Espinosa}}, \bibinfo {author} {\bibfnamefont {C.}~\bibnamefont {Vega}}, \
  and\ \bibinfo {author} {\bibfnamefont {E.}~\bibnamefont {Sanz}},\ }\href@noop
  {} {\bibfield  {journal} {\bibinfo  {journal} {The Journal of chemical
  physics}\ }\textbf {\bibinfo {volume} {141}},\ \bibinfo {pages} {134709}
  (\bibinfo {year} {2014})}\BibitemShut {NoStop}%
\bibitem [{\citenamefont {Espinosa}\ \emph {et~al.}(2015)\citenamefont
  {Espinosa}, \citenamefont {Vega}, \citenamefont {Valeriani},\ and\
  \citenamefont {Sanz}}]{espinosa2015crystalkk}%
  \BibitemOpen
  \bibfield  {author} {\bibinfo {author} {\bibfnamefont {J.~R.}\ \bibnamefont
  {Espinosa}}, \bibinfo {author} {\bibfnamefont {C.}~\bibnamefont {Vega}},
  \bibinfo {author} {\bibfnamefont {C.}~\bibnamefont {Valeriani}}, \ and\
  \bibinfo {author} {\bibfnamefont {E.}~\bibnamefont {Sanz}},\ }\href@noop {}
  {\bibfield  {journal} {\bibinfo  {journal} {The Journal of chemical physics}\
  }\textbf {\bibinfo {volume} {142}},\ \bibinfo {pages} {194709} (\bibinfo
  {year} {2015})}\BibitemShut {NoStop}%
\bibitem [{\citenamefont {Espinosa}\ \emph
  {et~al.}(2016{\natexlab{a}})\citenamefont {Espinosa}, \citenamefont {Vega},\
  and\ \citenamefont {Sanz}}]{espinosa2016icekk}%
  \BibitemOpen
  \bibfield  {author} {\bibinfo {author} {\bibfnamefont {J.~R.}\ \bibnamefont
  {Espinosa}}, \bibinfo {author} {\bibfnamefont {C.}~\bibnamefont {Vega}}, \
  and\ \bibinfo {author} {\bibfnamefont {E.}~\bibnamefont {Sanz}},\ }\href@noop
  {} {\bibfield  {journal} {\bibinfo  {journal} {The Journal of Physical
  Chemistry C}\ }\textbf {\bibinfo {volume} {120}},\ \bibinfo {pages} {8068}
  (\bibinfo {year} {2016}{\natexlab{a}})}\BibitemShut {NoStop}%
\bibitem [{\citenamefont {Sanchez-Burgos}\ \emph
  {et~al.}(2021{\natexlab{a}})\citenamefont {Sanchez-Burgos}, \citenamefont
  {Sanz}, \citenamefont {Vega},\ and\ \citenamefont
  {Espinosa}}]{sanchez2021fcckk}%
  \BibitemOpen
  \bibfield  {author} {\bibinfo {author} {\bibfnamefont {I.}~\bibnamefont
  {Sanchez-Burgos}}, \bibinfo {author} {\bibfnamefont {E.}~\bibnamefont
  {Sanz}}, \bibinfo {author} {\bibfnamefont {C.}~\bibnamefont {Vega}}, \ and\
  \bibinfo {author} {\bibfnamefont {J.~R.}\ \bibnamefont {Espinosa}},\
  }\href@noop {} {\bibfield  {journal} {\bibinfo  {journal} {Physical Chemistry
  Chemical Physics}\ }\textbf {\bibinfo {volume} {23}},\ \bibinfo {pages}
  {19611} (\bibinfo {year} {2021}{\natexlab{a}})}\BibitemShut {NoStop}%
\bibitem [{\citenamefont {Zeron}\ \emph {et~al.}(2022)\citenamefont {Zeron},
  \citenamefont {M{\'\i}guez}, \citenamefont {Mendiboure}, \citenamefont
  {Algaba},\ and\ \citenamefont {Blas}}]{zeron2022simulationkk}%
  \BibitemOpen
  \bibfield  {author} {\bibinfo {author} {\bibfnamefont {I.~M.}\ \bibnamefont
  {Zeron}}, \bibinfo {author} {\bibfnamefont {J.~M.}\ \bibnamefont
  {M{\'\i}guez}}, \bibinfo {author} {\bibfnamefont {B.}~\bibnamefont
  {Mendiboure}}, \bibinfo {author} {\bibfnamefont {J.}~\bibnamefont {Algaba}},
  \ and\ \bibinfo {author} {\bibfnamefont {F.~J.}\ \bibnamefont {Blas}},\
  }\href@noop {} {\bibfield  {journal} {\bibinfo  {journal} {The Journal of
  Chemical Physics}\ } (\bibinfo {year} {2022})}\BibitemShut {NoStop}%
\bibitem [{\citenamefont {Bekker}\ \emph {et~al.}(1993)\citenamefont {Bekker},
  \citenamefont {Berendsen}, \citenamefont {Dijkstra}, \citenamefont
  {Achterop}, \citenamefont {Vondrumen}, \citenamefont {VANDERSPOEL},
  \citenamefont {Sijbers}, \citenamefont {Keegstra},\ and\ \citenamefont
  {Renardus}}]{bekker1993gromacskk}%
  \BibitemOpen
  \bibfield  {author} {\bibinfo {author} {\bibfnamefont {H.}~\bibnamefont
  {Bekker}}, \bibinfo {author} {\bibfnamefont {H.}~\bibnamefont {Berendsen}},
  \bibinfo {author} {\bibfnamefont {E.}~\bibnamefont {Dijkstra}}, \bibinfo
  {author} {\bibfnamefont {S.}~\bibnamefont {Achterop}}, \bibinfo {author}
  {\bibfnamefont {R.}~\bibnamefont {Vondrumen}}, \bibinfo {author}
  {\bibfnamefont {D.}~\bibnamefont {VANDERSPOEL}}, \bibinfo {author}
  {\bibfnamefont {A.}~\bibnamefont {Sijbers}}, \bibinfo {author} {\bibfnamefont
  {H.}~\bibnamefont {Keegstra}}, \ and\ \bibinfo {author} {\bibfnamefont
  {M.}~\bibnamefont {Renardus}},\ }in\ \href@noop {} {\emph {\bibinfo
  {booktitle} {4th International Conference on Computational Physics (PC
  92)}}}\ (\bibinfo {organization} {World Scientific Publishing},\ \bibinfo
  {year} {1993})\ pp.\ \bibinfo {pages} {252--256}\BibitemShut {NoStop}%
\bibitem [{\citenamefont {Bussi}\ \emph {et~al.}(2007)\citenamefont {Bussi},
  \citenamefont {Donadio},\ and\ \citenamefont
  {Parrinello}}]{bussi2007canonicalkk}%
  \BibitemOpen
  \bibfield  {author} {\bibinfo {author} {\bibfnamefont {G.}~\bibnamefont
  {Bussi}}, \bibinfo {author} {\bibfnamefont {D.}~\bibnamefont {Donadio}}, \
  and\ \bibinfo {author} {\bibfnamefont {M.}~\bibnamefont {Parrinello}},\
  }\href@noop {} {\bibfield  {journal} {\bibinfo  {journal} {The Journal of
  chemical physics}\ }\textbf {\bibinfo {volume} {126}},\ \bibinfo {pages}
  {014101} (\bibinfo {year} {2007})}\BibitemShut {NoStop}%
\bibitem [{\citenamefont {Parrinello}\ and\ \citenamefont
  {Rahman}(1981)}]{parrinello1981polymorphickk}%
  \BibitemOpen
  \bibfield  {author} {\bibinfo {author} {\bibfnamefont {M.}~\bibnamefont
  {Parrinello}}\ and\ \bibinfo {author} {\bibfnamefont {A.}~\bibnamefont
  {Rahman}},\ }\href@noop {} {\bibfield  {journal} {\bibinfo  {journal}
  {Journal of Applied physics}\ }\textbf {\bibinfo {volume} {52}},\ \bibinfo
  {pages} {7182} (\bibinfo {year} {1981})}\BibitemShut {NoStop}%
\bibitem [{\citenamefont {Hockney}\ \emph {et~al.}(1974)\citenamefont
  {Hockney}, \citenamefont {Goel},\ and\ \citenamefont
  {Eastwood}}]{hockney1974quietkk}%
  \BibitemOpen
  \bibfield  {author} {\bibinfo {author} {\bibfnamefont {R.~W.}\ \bibnamefont
  {Hockney}}, \bibinfo {author} {\bibfnamefont {S.}~\bibnamefont {Goel}}, \
  and\ \bibinfo {author} {\bibfnamefont {J.}~\bibnamefont {Eastwood}},\
  }\href@noop {} {\bibfield  {journal} {\bibinfo  {journal} {Journal of
  Computational Physics}\ }\textbf {\bibinfo {volume} {14}},\ \bibinfo {pages}
  {148} (\bibinfo {year} {1974})}\BibitemShut {NoStop}%
\bibitem [{\citenamefont {Darden}\ \emph {et~al.}(1993)\citenamefont {Darden},
  \citenamefont {York},\ and\ \citenamefont {Pedersen}}]{darden1993particlekk}%
  \BibitemOpen
  \bibfield  {author} {\bibinfo {author} {\bibfnamefont {T.}~\bibnamefont
  {Darden}}, \bibinfo {author} {\bibfnamefont {D.}~\bibnamefont {York}}, \ and\
  \bibinfo {author} {\bibfnamefont {L.}~\bibnamefont {Pedersen}},\ }\href@noop
  {} {\bibfield  {journal} {\bibinfo  {journal} {The Journal of chemical
  physics}\ }\textbf {\bibinfo {volume} {98}},\ \bibinfo {pages} {10089}
  (\bibinfo {year} {1993})}\BibitemShut {NoStop}%
\bibitem [{\citenamefont {Essmann}\ \emph {et~al.}(1995)\citenamefont
  {Essmann}, \citenamefont {Perera}, \citenamefont {Berkowitz}, \citenamefont
  {Darden}, \citenamefont {Lee},\ and\ \citenamefont
  {Pedersen}}]{essmann1995smoothkk}%
  \BibitemOpen
  \bibfield  {author} {\bibinfo {author} {\bibfnamefont {U.}~\bibnamefont
  {Essmann}}, \bibinfo {author} {\bibfnamefont {L.}~\bibnamefont {Perera}},
  \bibinfo {author} {\bibfnamefont {M.~L.}\ \bibnamefont {Berkowitz}}, \bibinfo
  {author} {\bibfnamefont {T.}~\bibnamefont {Darden}}, \bibinfo {author}
  {\bibfnamefont {H.}~\bibnamefont {Lee}}, \ and\ \bibinfo {author}
  {\bibfnamefont {L.~G.}\ \bibnamefont {Pedersen}},\ }\href@noop {} {\bibfield
  {journal} {\bibinfo  {journal} {The Journal of chemical physics}\ }\textbf
  {\bibinfo {volume} {103}},\ \bibinfo {pages} {8577} (\bibinfo {year}
  {1995})}\BibitemShut {NoStop}%
\bibitem [{\citenamefont {Hess}\ \emph {et~al.}(1997)\citenamefont {Hess},
  \citenamefont {Bekker}, \citenamefont {Berendsen},\ and\ \citenamefont
  {Fraaije}}]{hess1997lincskk}%
  \BibitemOpen
  \bibfield  {author} {\bibinfo {author} {\bibfnamefont {B.}~\bibnamefont
  {Hess}}, \bibinfo {author} {\bibfnamefont {H.}~\bibnamefont {Bekker}},
  \bibinfo {author} {\bibfnamefont {H.~J.}\ \bibnamefont {Berendsen}}, \ and\
  \bibinfo {author} {\bibfnamefont {J.~G.}\ \bibnamefont {Fraaije}},\
  }\href@noop {} {\bibfield  {journal} {\bibinfo  {journal} {Journal of
  computational chemistry}\ }\textbf {\bibinfo {volume} {18}},\ \bibinfo
  {pages} {1463} (\bibinfo {year} {1997})}\BibitemShut {NoStop}%
\bibitem [{\citenamefont {Lamas}\ \emph {et~al.}(2021)\citenamefont {Lamas},
  \citenamefont {Espinosa}, \citenamefont {Conde}, \citenamefont
  {Ram{\'\i}rez}, \citenamefont {de~Hijes}, \citenamefont {Noya}, \citenamefont
  {Vega},\ and\ \citenamefont {Sanz}}]{lamas2021homogeneouskk}%
  \BibitemOpen
  \bibfield  {author} {\bibinfo {author} {\bibfnamefont {C.}~\bibnamefont
  {Lamas}}, \bibinfo {author} {\bibfnamefont {J.}~\bibnamefont {Espinosa}},
  \bibinfo {author} {\bibfnamefont {M.}~\bibnamefont {Conde}}, \bibinfo
  {author} {\bibfnamefont {J.}~\bibnamefont {Ram{\'\i}rez}}, \bibinfo {author}
  {\bibfnamefont {P.~M.}\ \bibnamefont {de~Hijes}}, \bibinfo {author}
  {\bibfnamefont {E.~G.}\ \bibnamefont {Noya}}, \bibinfo {author}
  {\bibfnamefont {C.}~\bibnamefont {Vega}}, \ and\ \bibinfo {author}
  {\bibfnamefont {E.}~\bibnamefont {Sanz}},\ }\href@noop {} {\bibfield
  {journal} {\bibinfo  {journal} {Physical Chemistry Chemical Physics}\
  }\textbf {\bibinfo {volume} {23}},\ \bibinfo {pages} {26843} (\bibinfo {year}
  {2021})}\BibitemShut {NoStop}%
\bibitem [{\citenamefont {Espinosa}\ \emph
  {et~al.}(2016{\natexlab{b}})\citenamefont {Espinosa}, \citenamefont {Young},
  \citenamefont {Jiang}, \citenamefont {Gupta}, \citenamefont {Vega},
  \citenamefont {Sanz}, \citenamefont {Debenedetti},\ and\ \citenamefont
  {Panagiotopoulos}}]{espinosa2016calculationkk}%
  \BibitemOpen
  \bibfield  {author} {\bibinfo {author} {\bibfnamefont {J.}~\bibnamefont
  {Espinosa}}, \bibinfo {author} {\bibfnamefont {J.}~\bibnamefont {Young}},
  \bibinfo {author} {\bibfnamefont {H.}~\bibnamefont {Jiang}}, \bibinfo
  {author} {\bibfnamefont {D.}~\bibnamefont {Gupta}}, \bibinfo {author}
  {\bibfnamefont {C.}~\bibnamefont {Vega}}, \bibinfo {author} {\bibfnamefont
  {E.}~\bibnamefont {Sanz}}, \bibinfo {author} {\bibfnamefont {P.~G.}\
  \bibnamefont {Debenedetti}}, \ and\ \bibinfo {author} {\bibfnamefont {A.~Z.}\
  \bibnamefont {Panagiotopoulos}},\ }\href@noop {} {\bibfield  {journal}
  {\bibinfo  {journal} {The Journal of chemical physics}\ }\textbf {\bibinfo
  {volume} {145}},\ \bibinfo {pages} {154111} (\bibinfo {year}
  {2016}{\natexlab{b}})}\BibitemShut {NoStop}%
\bibitem [{\citenamefont {Jiang}\ \emph {et~al.}(2018)\citenamefont {Jiang},
  \citenamefont {Haji-Akbari}, \citenamefont {Debenedetti},\ and\ \citenamefont
  {Panagiotopoulos}}]{jiang2018forwardkk}%
  \BibitemOpen
  \bibfield  {author} {\bibinfo {author} {\bibfnamefont {H.}~\bibnamefont
  {Jiang}}, \bibinfo {author} {\bibfnamefont {A.}~\bibnamefont {Haji-Akbari}},
  \bibinfo {author} {\bibfnamefont {P.~G.}\ \bibnamefont {Debenedetti}}, \ and\
  \bibinfo {author} {\bibfnamefont {A.~Z.}\ \bibnamefont {Panagiotopoulos}},\
  }\href@noop {} {\bibfield  {journal} {\bibinfo  {journal} {The Journal of
  chemical physics}\ }\textbf {\bibinfo {volume} {148}},\ \bibinfo {pages}
  {044505} (\bibinfo {year} {2018})}\BibitemShut {NoStop}%
\bibitem [{\citenamefont {Zimmermann}\ \emph {et~al.}(2018)\citenamefont
  {Zimmermann}, \citenamefont {Vorselaars}, \citenamefont {Espinosa},
  \citenamefont {Quigley}, \citenamefont {Smith}, \citenamefont {Sanz},
  \citenamefont {Vega},\ and\ \citenamefont {Peters}}]{zimmermann2018naclkk}%
  \BibitemOpen
  \bibfield  {author} {\bibinfo {author} {\bibfnamefont {N.~E.}\ \bibnamefont
  {Zimmermann}}, \bibinfo {author} {\bibfnamefont {B.}~\bibnamefont
  {Vorselaars}}, \bibinfo {author} {\bibfnamefont {J.~R.}\ \bibnamefont
  {Espinosa}}, \bibinfo {author} {\bibfnamefont {D.}~\bibnamefont {Quigley}},
  \bibinfo {author} {\bibfnamefont {W.~R.}\ \bibnamefont {Smith}}, \bibinfo
  {author} {\bibfnamefont {E.}~\bibnamefont {Sanz}}, \bibinfo {author}
  {\bibfnamefont {C.}~\bibnamefont {Vega}}, \ and\ \bibinfo {author}
  {\bibfnamefont {B.}~\bibnamefont {Peters}},\ }\href@noop {} {\bibfield
  {journal} {\bibinfo  {journal} {The Journal of chemical physics}\ }\textbf
  {\bibinfo {volume} {148}},\ \bibinfo {pages} {222838} (\bibinfo {year}
  {2018})}\BibitemShut {NoStop}%
\bibitem [{\citenamefont {Lechner}\ and\ \citenamefont
  {Dellago}(2008)}]{lechner2008accuratekk}%
  \BibitemOpen
  \bibfield  {author} {\bibinfo {author} {\bibfnamefont {W.}~\bibnamefont
  {Lechner}}\ and\ \bibinfo {author} {\bibfnamefont {C.}~\bibnamefont
  {Dellago}},\ }\href@noop {} {\bibfield  {journal} {\bibinfo  {journal} {The
  Journal of chemical physics}\ }\textbf {\bibinfo {volume} {129}},\ \bibinfo
  {pages} {114707} (\bibinfo {year} {2008})}\BibitemShut {NoStop}%
\bibitem [{\citenamefont {Espinosa}\ \emph
  {et~al.}(2016{\natexlab{c}})\citenamefont {Espinosa}, \citenamefont {Vega},
  \citenamefont {Valeriani},\ and\ \citenamefont {Sanz}}]{espinosa2016seedingkk}%
  \BibitemOpen
  \bibfield  {author} {\bibinfo {author} {\bibfnamefont {J.~R.}\ \bibnamefont
  {Espinosa}}, \bibinfo {author} {\bibfnamefont {C.}~\bibnamefont {Vega}},
  \bibinfo {author} {\bibfnamefont {C.}~\bibnamefont {Valeriani}}, \ and\
  \bibinfo {author} {\bibfnamefont {E.}~\bibnamefont {Sanz}},\ }\href@noop {}
  {\bibfield  {journal} {\bibinfo  {journal} {The Journal of chemical physics}\
  }\textbf {\bibinfo {volume} {144}},\ \bibinfo {pages} {034501} (\bibinfo
  {year} {2016}{\natexlab{c}})}\BibitemShut {NoStop}%
\bibitem [{\citenamefont {Sanchez-Burgos}\ \emph {et~al.}(2022)\citenamefont
  {Sanchez-Burgos}, \citenamefont {Tejedor}, \citenamefont {Vega},
  \citenamefont {Conde}, \citenamefont {Sanz}, \citenamefont {Ramirez},\ and\
  \citenamefont {Espinosa}}]{sanchez2022homogeneouskk}%
  \BibitemOpen
  \bibfield  {author} {\bibinfo {author} {\bibfnamefont {I.}~\bibnamefont
  {Sanchez-Burgos}}, \bibinfo {author} {\bibfnamefont {A.~R.}\ \bibnamefont
  {Tejedor}}, \bibinfo {author} {\bibfnamefont {C.}~\bibnamefont {Vega}},
  \bibinfo {author} {\bibfnamefont {M.~M.}\ \bibnamefont {Conde}}, \bibinfo
  {author} {\bibfnamefont {E.}~\bibnamefont {Sanz}}, \bibinfo {author}
  {\bibfnamefont {J.}~\bibnamefont {Ramirez}}, \ and\ \bibinfo {author}
  {\bibfnamefont {J.~R.}\ \bibnamefont {Espinosa}},\ }\href@noop {} {\bibfield
  {journal} {\bibinfo  {journal} {The Journal of Chemical Physics}\ }\textbf
  {\bibinfo {volume} {157}},\ \bibinfo {pages} {094503} (\bibinfo {year}
  {2022})}\BibitemShut {NoStop}%
\bibitem [{\citenamefont {Sanchez-Burgos}\ \emph
  {et~al.}(2021{\natexlab{b}})\citenamefont {Sanchez-Burgos}, \citenamefont
  {Garaizar}, \citenamefont {Vega}, \citenamefont {Sanz},\ and\ \citenamefont
  {Espinosa}}]{sanchez2021parasitickk}%
  \BibitemOpen
  \bibfield  {author} {\bibinfo {author} {\bibfnamefont {I.}~\bibnamefont
  {Sanchez-Burgos}}, \bibinfo {author} {\bibfnamefont {A.}~\bibnamefont
  {Garaizar}}, \bibinfo {author} {\bibfnamefont {C.}~\bibnamefont {Vega}},
  \bibinfo {author} {\bibfnamefont {E.}~\bibnamefont {Sanz}}, \ and\ \bibinfo
  {author} {\bibfnamefont {J.~R.}\ \bibnamefont {Espinosa}},\ }\href@noop {}
  {\bibfield  {journal} {\bibinfo  {journal} {Soft Matter}\ }\textbf {\bibinfo
  {volume} {17}},\ \bibinfo {pages} {489} (\bibinfo {year}
  {2021}{\natexlab{b}})}\BibitemShut {NoStop}%
\bibitem [{\citenamefont {Wulff}(1901)}]{wulff1901xxvkk}%
  \BibitemOpen
  \bibfield  {author} {\bibinfo {author} {\bibfnamefont {G.}~\bibnamefont
  {Wulff}},\ }\href@noop {} {\bibfield  {journal} {\bibinfo  {journal}
  {Zeitschrift f{\"u}r Kristallographie-Crystalline Materials}\ }\textbf
  {\bibinfo {volume} {34}},\ \bibinfo {pages} {449} (\bibinfo {year}
  {1901})}\BibitemShut {NoStop}%
\bibitem [{\citenamefont {Rahm}\ and\ \citenamefont
  {Erhart}(2020)}]{rahm2020wulffpackkk}%
  \BibitemOpen
  \bibfield  {author} {\bibinfo {author} {\bibfnamefont {J.~M.}\ \bibnamefont
  {Rahm}}\ and\ \bibinfo {author} {\bibfnamefont {P.}~\bibnamefont {Erhart}},\
  }\href@noop {} {\bibfield  {journal} {\bibinfo  {journal} {Journal of Open
  Source Software}\ }\textbf {\bibinfo {volume} {5}},\ \bibinfo {pages} {1944}
  (\bibinfo {year} {2020})}\BibitemShut {NoStop}%
\end{thebibliography}
\end{document}